%% file: main.tex
\documentclass[acmsmall,screen,nonacm]{acmart}

\setcopyright{none}

\input{packages}
\input{macros}

\input{generated/bug_numbers}
\input{generated/rq2_numbers}

\begin{document}

\pagenumbering{arabic}

\title{A Multi-Domain Study on the Effectiveness of LLMs at Writing Property-Based Tests}

\author{M M Abid Naziri}\authornote{Equal Contribution.}
\orcid{0000-0002-3499-5283}
\affiliation{%
  \institution{NC State University}
  \city{Raleigh}
  \state{NC}
  \country{USA}
}
\email{mnaziri@ncsu.edu}

\author{Feiran (Alex) Qin}\authornotemark[1]
\orcid{0009-0003-9310-0798}
\affiliation{%
  \institution{NC State University}
  \city{Raleigh}
  \state{NC}
  \country{USA}
}
\email{fqin2@ncsu.edu}

\author{Khush Patel}
\orcid{0009-0003-9484-3753}
\affiliation{%
  \institution{NC State University}
  \city{Raleigh}
  \state{NC}
  \country{USA}
}
\email{kpatel53@ncsu.edu}

\author{Marcelo d'Amorim}
\orcid{0000-0002-1323-8769}
\affiliation{%
  \institution{NC State University}
  \city{Raleigh}
  \state{NC}
  \country{USA}
}
\email{mdamori@ncsu.edu}

\begin{abstract}
A property-based test (PBT) checks a property of a program on many generated
inputs, but writing one takes a property, a generator of valid inputs, and an
oracle. LLM-based PBT writers produce all three, yet they have been evaluated
only on Python libraries. We study how two of them, \proptestai and
\agenticpbt, generalize to \numPrograms{} programs in \numDomains{} domains:
deep learning libraries, databases, system software, compilers, and LLM
serving engines. Most of these programs take text in their own language, such
as queries, configurations, and programs. \agenticpbt covers more branches
than \proptestai on seven of the ten programs with Opus~5 and on all
ten with GLM-5.3-Flash.
Domain-specific fuzzers cover more of the compilers and of PostgreSQL and less
of PyTorch than \agenticpbt. We filed \bugsReported{} bug reports, \bugsDuplicate{} of which duplicate
reports that others filed after the models' knowledge cutoffs; maintainers
have confirmed \keynum{\the\numexpr\bugsConfirmed-\bugsDuplicate\relax{}} of
the others, in all five domains, and fixed \bugsFixed{}. We draw lessons for building and evaluating such writers: low observability
hurts them, and mature fuzzers paired with LLM-written properties could check
that each feature behaves as documented.
\end{abstract}

\maketitle
\renewcommand{\shortauthors}{Naziri et al.}
\newcounter{mylessoncounter}
\input{intro.tex}
\input{background.tex}
\input{domains.tex}

\input{eval.tex}

\input{discussion.tex}
\input{related.tex}

\input{conclusion.tex}
\input{data.tex}

\bibliographystyle{ACM-Reference-Format}
\bibliography{main} 

\balance

\end{document}

%% file: packages.tex
\usepackage{amsmath}
\usepackage{color, soul, xcolor}  %
\usepackage{colortbl}
\usepackage{booktabs} %
\usepackage{enumitem}
\usepackage{caption}
\usepackage{amsmath}
\usepackage{listings}
\usepackage[frozencache,cachedir=minted-cache]{minted}
\usepackage{balance}
\usepackage{makecell}
\usepackage[normalem]{ulem} %
\usepackage{multirow}
\usepackage{tabularx}
\usepackage{longtable}
\usepackage{romannum}
\usepackage{textcomp}
\usepackage[T1]{fontenc}
\usepackage{newfloat}

\usepackage{tikz}
\usepackage{pgfplots}
\pgfplotsset{compat=1.18}
\usepackage{fontawesome5}
\tikzset{>=latex}
\usetikzlibrary{calc}
\usetikzlibrary{shapes.geometric}
\usetikzlibrary{decorations.pathreplacing}
\usetikzlibrary{positioning}

\usetikzlibrary{positioning}
\usetikzlibrary{tikzmark}
\usepackage[htt]{hyphenat}
\usepackage{float}
\definecolor{shadecolor}{gray}{0.97}
\usepackage[most]{tcolorbox}
\tcbuselibrary{minted}
\usepackage[linewidth=1pt]{mdframed}
\usepackage{array}
\usepackage{CJKutf8}
\usepackage{pifont}  %
\usepackage{subcaption}
\usepackage{xspace}
\usepackage{wrapfig}
\usepackage{needspace}  %
\usepackage{pythonhighlight}
\usepackage{algorithm}
\usepackage{algpseudocode}
\lstdefinestyle{pseudocode-style}{
  basicstyle=\scriptsize\ttfamily,
  keywordstyle=\scriptsize\ttfamily\bfseries,
  language=Python,             %
  frame=none,              %
  escapechar=|,
  aboveskip=2pt,
  belowskip=2pt,
  breaklines=true,           %
  breakatwhitespace=true,   %
  showspaces=false,
  numbers=left,
  numbersep=4pt,              %
  tabsize=4,                %
  extendedchars=true,         %
  keywords=[2]{tcp, flag, threshold, track, count, seconds, classtype, sid, msg, flow, max, size, abs}
}
\lstnewenvironment{pseudocode}[1][]{%
	\lstset{style=pseudocode-style, #1}%
}{}
\usepackage{placeins}  %

%% file: macros.tex
\newcolumntype{C}[1]{>{\centering\arraybackslash}p{#1}}

\newcommand{\Code}[1]{{\small\ifmmode{\texttt{#1}}\else$\texttt{#1}$\fi}}
\newcommand{\CodeIn}[1]{{\small\ifmmode{\mathtt{#1}}\else$\mathtt{#1}$\fi}}

\definecolor{mintbg}{rgb}{0.94, 1.0, 1.0} %

\newcommand{\MYROMAN}[1]{%
  \textup{\uppercase\expandafter{\romannumeral#1}}%
}
\newcommand{\Sum}[2]{\the\numexpr #1 + #2 \relax}

\newcommand{\RepoURL}{\url{https://anonymous.4open.science/r/pbt-study-artifact-6A0B}}
\newcommand{\eg}{e.g.}

\newcommand{\numDomains}{five}
\newcommand{\numProgramsPerDomain}{two}

\newcommand{\numPrograms}{ten}

\newcommand{\rqtwoTests}{2,622}
\newcommand{\rqtwoSample}{500}

\newcommand{\proptestai}{\textsc{Proptest-AI}\xspace}
\newcommand{\agenticpbt}{\textsc{Agentic-PBT}\xspace}

\newcommand{\lessonblock}[2]{
\stepcounter{mylessoncounter}
\begin{tcolorbox}[
colback=black!10!white, colframe=black, boxsep=0pt, left=4pt, right=4pt,
top=2pt, bottom=2pt, arc=1pt, boxrule=0.6pt, before skip=5pt, after skip=5pt]
\textbf{Answer to #1:} #2
\end{tcolorbox}
}

\usepackage[normalem]{ulem} %

\newenvironment{packed_itemize}{
\vspace{-0.1ex}
\begin{list}{\labelitemi}{\setlength{\leftmargin}{1.2em}\setlength{\labelwidth}{0.8em}\setlength{\labelsep}{0.4em}\setlength{\itemindent}{0pt}\setlength{\listparindent}{0pt}}
 \setlength{\itemsep}{0pt} \setlength{\parskip}{0pt} \setlength{\parsep}{0pt} \setlength{\headsep}{0pt} \setlength{\topskip}{0pt} \setlength{\topmargin}{0pt} \setlength{\topsep}{0pt} \setlength{\partopsep}{0pt}
 }{\end{list}
\vspace{-0.1ex}
}

\tcbset{codebox/.style={listing engine=minted, listing only,
  colback=white, colframe=black, boxrule=0.4pt, arc=0pt, outer arc=0pt,
  left=2pt, right=2pt, top=1pt, bottom=1pt, before skip=0pt, after skip=2pt}}
\tcbset{partbox/.style={codebox, fonttitle=\scriptsize\bfseries, coltitle=black,
  colbacktitle=black!8, toptitle=0pt, bottomtitle=0pt, lefttitle=2pt}}
\newcommand{\codehl}[1]{{\setlength{\fboxsep}{0.5pt}\colorbox{black!12}{#1}}}
\newcommand{\promptslot}[1]{{\setlength{\fboxsep}{0.8pt}\colorbox{black!12}{\strut#1}}}
\newcommand{\promptraw}[1]{\promptslot{\texttt{\footnotesize #1}}}
\providecommand{\codehl}[1]{\promptslot{#1}}
\newcommand{\promptparam}[1]{\promptraw{\{#1\}}}

\newif\ifshowdraft
\showdraftfalse
\newcommand{\draftnum}[1]{\ifshowdraft\textcolor{draftcolor}{#1}\else#1\fi}
\newcommand{\drafttext}[1]{\ifshowdraft\textcolor{draftcolor}{#1}\else#1\fi}
\newcommand{\keynum}[1]{\textbf{#1}}
\definecolor{draftcolor}{HTML}{8B4513}

\newcommand{\lessonmark}[1]{{\setlength{\ULdepth}{1.2pt}\uline{#1}}}

\makeatletter
\renewcommand\section{\@startsection{section}{1}{\z@}%
  {-.5\baselineskip \@plus -1\p@ \@minus -.2\p@}%
  {.15\baselineskip}%
  {\ACM@NRadjust\@secfont}}
\renewcommand\subsection{\@startsection{subsection}{2}{\z@}%
  {-.5\baselineskip \@plus -1\p@ \@minus -.2\p@}%
  {.15\baselineskip}%
  {\ACM@NRadjust\@subsecfont}}
\renewcommand\subsubsection{\@startsection{subsubsection}{3}{\z@}%
  {-.25\baselineskip \@plus -1\p@ \@minus -.2\p@}%
  {-3.5\p@}%
  {\ACM@NRadjust{\@subsubsecfont\@adddotafter}}}
\makeatother

%% file: generated/bug_numbers.tex
\newcommand{\bugsSystems}{10}                   %
\newcommand{\bugsReportedRQone}{30}             %
\newcommand{\bugsFocal}{742}                    %
\newcommand{\bugsReported}{180}                 %
\newcommand{\bugsConfirmed}{109}                %
\newcommand{\bugsRejected}{3}                   %
\newcommand{\bugsPending}{68}                   %
\newcommand{\bugsFixed}{23}                     %
\newcommand{\bugsDuplicate}{44}                 %
\newcommand{\bugsPerFocal}{0.15}                %
\newcommand{\bugsPerFocalNew}{0.09}             %
\newcommand{\bugsPerFocalReported}{0.24}        %
\newcommand{\bugsRejectRate}{2}                 %

\newcommand{\bugsReportedFocal}{180}            %
\newcommand{\bugsConfirmedFocal}{109}           %
\newcommand{\bugsConfirmedNewFocal}{65}         %
\newcommand{\bugsCrash}{24}                     %
\newcommand{\bugsFunctional}{156}               %

\newcommand{\bugsFocalDL}{180}                  %
\newcommand{\bugsReportedDL}{50}                %
\newcommand{\bugsRejectedDL}{1}                 %
\newcommand{\bugsRejectRateDL}{2}               %
\newcommand{\bugsFocalServing}{153}             %
\newcommand{\bugsReportedServing}{66}           %
\newcommand{\bugsConfirmedServing}{44}          %
\newcommand{\bugsRejectedServing}{1}            %
\newcommand{\bugsDuplicateServing}{36}          %
\newcommand{\bugsConfirmedNewFocalServing}{8}   %
\newcommand{\bugsPerFocalNewServing}{0.05}      %
\newcommand{\bugsRejectRateServing}{2}          %
\newcommand{\bugsFocalDatabase}{240}            %
\newcommand{\bugsReportedDatabase}{32}          %
\newcommand{\bugsPendingDatabase}{26}           %
\newcommand{\bugsConfirmedNewFocalDatabase}{5}  %
\newcommand{\bugsPerFocalNewDatabase}{0.02}     %
\newcommand{\bugsPerFocalMaxDatabase}{0.13}     %
\newcommand{\bugsFocalSystem}{77}               %
\newcommand{\bugsConfirmedNewFocalSystem}{14}   %
\newcommand{\bugsPerFocalNewSystem}{0.18}       %
\newcommand{\bugsFocalCompiler}{92}             %
\newcommand{\bugsReportedCompiler}{14}          %
\newcommand{\bugsRejectedCompiler}{1}           %
\newcommand{\bugsRejectRateCompiler}{7}         %

\newcommand{\bugsReportedPostgreSQL}{9}         %
\newcommand{\bugsPendingPostgreSQL}{8}          %

%% file: generated/rq2_numbers.tex
\newcommand{\rqtwoImplicitTotal}{1.5}           %
\newcommand{\rqtwoContractLLM}{72}              %
\newcommand{\rqtwoContractDB}{40}               %
\newcommand{\rqtwoContractTotal}{54}            %
\newcommand{\rqtwoAlgebraicTotal}{3.5}          %
\newcommand{\rqtwoDifferentialDL}{27}           %
\newcommand{\rqtwoDifferentialTotal}{19.3}      %
\newcommand{\rqtwoMetamorphicSYS}{24}           %
\newcommand{\rqtwoMetamorphicTotal}{21.3}       %

\newcommand{\rqtwoCompareDL}{51}                %
\newcommand{\rqtwoCompareLLM}{22}               %
\newcommand{\rqtwoCompareDB}{51}                %
\newcommand{\rqtwoCompareTotal}{41}             %
\newcommand{\rqtwoNonContractTotal}{46}         %

\newcommand{\rqtwoContractInvariant}{47.5}      %
\newcommand{\rqtwoContractRejection}{6.1}       %
\newcommand{\rqtwoContractDocumented}{0.8}      %

\newcommand{\rqtwoSourceNoneTotal}{5}           %
\newcommand{\rqtwoSourceDocumentationDL}{78}    %
\newcommand{\rqtwoSourceDocumentationLLM}{56}   %
\newcommand{\rqtwoSourceDocumentationDB}{89}    %
\newcommand{\rqtwoSourceDocumentationTotal}{78} %
\newcommand{\rqtwoSourceCodeLLM}{38}            %
\newcommand{\rqtwoSourceCodeSYS}{14}            %
\newcommand{\rqtwoSourceCodeCC}{8}              %
\newcommand{\rqtwoSourceCodeTotal}{14}          %

\newcommand{\rqtwoDocstring}{41}                %
\newcommand{\rqtwoDocFile}{37}                  %

\newcommand{\rqtwoNoSourceNotFound}{3.2}        %
\newcommand{\rqtwoNoSourceGeneral}{0.9}         %
\newcommand{\rqtwoNoSourceImplicit}{0.7}        %

\newcommand{\rqtwoCIDomain}{12.4}               %
\newcommand{\rqtwoCITotal}{6.0}                 %
\newcommand{\rqtwoCIContrast}{17}               %

\newcommand{\rqtwoWithSource}{478}              %
\newcommand{\rqtwoNoSource}{22}                 %

%% file: intro.tex
\section{Introduction}
\label{sec:intro}

A property is a predicate over software behavior that must hold for every input
in a declared input space~\cite{dwyer1999patterns,claessen2000quickcheck,Hieron_ETAL_CSUR_2009}.
A property-based test (PBT)~\cite{claessen2000quickcheck} checks a property on
many generated inputs: a \emph{generator} draws inputs from the input space,
and an \emph{oracle} checks the property on each input and the program's
output. PBT libraries such as QuickCheck~\cite{claessen2000quickcheck} and
Hypothesis~\cite{maciver2019hypothesis} run the test and automatically reduce a failing input
into a small counterexample. They do not write the test itself. Writing
a PBT takes a property worth checking, a generator that respects the
constraints the program places on its inputs, and an oracle that knows what the
program must compute.

Properties can be mined from a project's artifacts~\cite{barr2015oracle}, such
as its tests, documentation, and commit history. Techniques for mining
properties go back more than two decades~\cite{lo2011specification,
robillard2013automated, germiniani2025mining, ernst2001dynamically,
ammons2002mining, yang2006perracotta, wasylkowski2007detecting,
lorenzoli2008automatic, gabel2008javert, legoues2009specification,
nguyen2009graphbased}, and many tools turn mined properties into tests that
find bugs~\cite{goffi2016toradocu, peleg2018jarvis, tsukamoto2018autoput,
kang2021adversarial, blasi2021memo, Xie_ETAL_ISSTA22, blasi2022callmemaybe,
tiwari2024proze, xu2024mrscout, glock2025teralizer}. For example, Toradocu~\cite{goffi2016toradocu} mines API documentation,
MR-Scout~\cite{xu2024mrscout} existing tests, and
DeepREL~\cite{deng2022fuzzing} the signatures of deep learning APIs. Each of these tools is built for
one \emph{source}, the kind of artifact it mines, and one \emph{domain}, a
class of programs such as deep learning libraries. Moving it to a new source or domain means rebuilding its mining logic and
generator.

Large Language Models (LLMs) can read any kind of software artifact, which
would replace the per-source mining logic, and they can write the generator
and the oracle that a new program needs. We call a tool that uses an LLM to write PBTs an \emph{LLM-based PBT writer}.
Two general-purpose ones have been proposed. \proptestai~\cite{vikram2024pbt} prompts an LLM with the documentation
of one Python API method. \agenticpbt~\cite{maaz2025agentic} prompts a coding agent, an LLM that can run commands and edit files, to read
a Python module's code, write PBTs, run them, revise a test whose failure it
judges a false alarm rather than a bug, and report the remaining failures as
bugs. Both tools were evaluated only on Python libraries,
which take Python values as inputs and need no \emph{harness}, the code that
starts or feeds the program under test.

Their evaluations therefore leave open whether an LLM can write PBTs for
programs whose inputs are text in the program's own language, such as
queries, configuration files, or programs. Such an input is valid only under
the program's semantics, and a test usually reaches the program through a
process or a server. For such programs, a developer must write a
generator that builds valid inputs, and an oracle that either predicts the
result or checks it against a second computation. A PBT writer that generalizes would bring property-based
testing to database servers, compilers, and system software. Compilers and
databases already have mature testing tools~\cite{yang2011csmith, le2014emi,
livinskii2020yarpgen, ou2024metamut, slutz1998rags, rigger2020pqs,
rigger2020norec, rigger2020tlp, zhong2020squirrel}, and such a writer would complement them with one tool that needs only a
small harness per domain. We therefore ask:
\emph{how well does an LLM-based PBT writer generalize across domains, and
where does it fall short?} To our knowledge, no study has evaluated such a writer on these programs or
compared its tests across domains.

\textbf{Our Work.} We answer this question with a study of \numDomains{}
domains: deep learning libraries, databases, system software, compilers, and
LLM serving engines. We chose widely used domains that differ in how a test
reaches the program and in the inputs it must build
(Section~\ref{sec:domains}). We study \numProgramsPerDomain{} programs per
domain, \numPrograms{} programs in total: JAX~\cite{jax2018github} and
PyTorch~\cite{paszke2019pytorch} (deep learning libraries),
PostgreSQL~\cite{postgresql} and ArangoDB~\cite{arangodb} (databases),
OpenSSH~\cite{openssh} and cronie~\cite{cronie} (system software),
GCC~\cite{gcc} and Clang~\cite{clang} (compilers), and vLLM~\cite{kwon2023vllm}
and SGLang~\cite{zheng2024sglang} (LLM serving engines). Databases, system
software, and compilers are reached only through a process or a server, which
no prior evaluation covered. For these domains, we give the tools a harness that only runs the program and returns its output, \eg{}, a
module that starts a database server and returns the rows of a query. We write one harness per domain, in about 30 minutes with an LLM's help, and
the tests of its two programs share it. The
harness does not check the output. The LLM writes all generators, properties,
and oracles. Deep learning libraries serve as a control: a test calls them
directly with Python values and needs no harness, as in the prior evaluations. LLM serving needs no harness either: tests call the engines' Python APIs. Besides the harness, we change only the 32 of \agenticpbt's 290 prompt lines
that locate and read the target, and none of \proptestai's prompts. Each test targets a \emph{focal point}, one feature of a program, such as an SQL function or a compiler builtin.

The study has four research questions. RQ1 assesses \emph{performance} and
selects the writer that RQ2 to RQ4 study. We compare the two LLM-based PBT
writers with each other on all ten programs, with two LLMs from different
vendors, Claude Opus~5 and GLM-5.3-Flash. We measure branch coverage and
examine failure reports. RQ2 studies the selected writer, \agenticpbt,
per domain: the oracles of its tests and the sources of their properties.
RQ3 counts the \emph{confirmed bugs} that the tests reveal, those that
maintainers accept as real. Because an LLM may have seen a bug report during
training, a previously reported bug counts only if its first public report
postdates the knowledge cutoff of the model that found it.
RQ4 compares \agenticpbt with domain-specific tools in three domains, in
branch coverage and in the crashes that each reveals: Centaur~\cite{naziri2026centaur} and
FlashFuzz~\cite{qin2026effective} on PyTorch, MetaMut~\cite{ou2024metamut}
and GrayC~\cite{evenmendoza2023grayc} on GCC and Clang, and
SQLancer~\cite{rigger2020tlp} on PostgreSQL.

\textbf{Results.} \agenticpbt covers more branches than \proptestai on seven of the
ten programs with Opus~5 and on all ten with GLM-5.3-Flash.
Over 12 hours, the domain-specific tools cover more of the compilers and of
PostgreSQL, where \agenticpbt reaches \draftnum{73\%} to \draftnum{85\%} of
the leading tool's coverage, and \agenticpbt covers \draftnum{1.1} to
\draftnum{1.4} times as much of PyTorch as FlashFuzz and Centaur. On the compilers, MetaMut and GrayC find \draftnum{23} and
\draftnum{4} crashes and \agenticpbt \draftnum{4}, none in common.
The documentation states the property of \rqtwoSourceDocumentationTotal\% of
the tests, and \rqtwoCompareTotal\% of their oracles compare the output with
another run or another implementation rather than check a stated condition.
We filed \bugsReported{} bug reports from \agenticpbt's tests. Among them are
\bugsDuplicate{} \emph{duplicates}: bugs that others reported independently after
the knowledge cutoff of the model that found them. Of the other
\the\numexpr\bugsReported-\bugsDuplicate\relax{} reports, maintainers have
confirmed \keynum{\the\numexpr\bugsConfirmed-\bugsDuplicate\relax{}}, in every
domain and all ten programs, fixed \bugsFixed{}, and rejected \bugsRejected{};
the rest await a response. Most duplicates are in LLM serving,
\bugsDuplicateServing{} of the \bugsDuplicate{}.

\textbf{Key Lessons.} We highlight two of the lessons of Section~\ref{sec:lessons}.
\lessonmark{First}, low observability hurts PBT writers. A property can be
documented and still be hard to check when the program does not show the
behavior that the property concerns, which affects system-level tests most.
A cronie test, for example, has to exceed one of cronie's limits to reveal how
cronie reads a crontab line. \lessonmark{Second}, mature fuzzers and LLM-based writers contribute different
parts of a test. The fuzzers reach more of the compilers and of PostgreSQL,
but they check crashes or, in SQLancer's case, one relation between queries.
The writers take most of their properties from the documentation, and most of the bugs \agenticpbt found are wrong results that only such a
property detects.
Future work could have MetaMut or SQLancer check properties that an LLM
extracts from the documentation.

We make three contributions: (1)~a study of general-purpose LLM-based PBT
writers on \numPrograms{} programs in \numDomains{} domains, most of which
take text in their own language, which measures the coverage, failures,
oracles, and bugs of their tests (Section~\ref{sec:eval}); (2)~lessons for
building and evaluating such writers (Section~\ref{sec:lessons}); and
(3)~\keynum{\the\numexpr\bugsConfirmed-\bugsDuplicate\relax{}} new bugs
confirmed by maintainers, and an artifact with the harnesses, prompts,
scripts, tests, and bug reports of the study, so that others can repeat it
with other models and writers (Data Availability).

%% file: background.tex
\section{Background}
\label{sec:background}

\subsection{Property-Based Testing}

\label{sec:bg-pbt}

\textbf{Definition.} Property-based testing checks that a program
satisfies a property on many generated inputs~\cite{claessen2000quickcheck}.
A program $f: X \to Y$ usually accepts only part of its input space $X$, the
inputs that satisfy its precondition $\mathit{pre}$; we call them
\emph{valid}. Given a property $P$, a property-based test approximates
$\forall x \in X : \mathit{pre}(x) \Rightarrow P(x, f(x))$ by checking it on
sampled inputs, because $X$ is usually too large to enumerate. A property-based test (PBT) has two parts. A
\emph{generator} samples inputs $x$ from $X$; it should draw only valid ones. An \emph{oracle} evaluates $P(x,
f(x))$ on each sampled input $x$ and the program's output $f(x)$, usually as an
assertion, and decides whether the execution is
acceptable~\cite{barr2015oracle}. The test fails when the oracle rejects an
input, and that input is a counterexample to the property. A failure on an
invalid input, or on a property the program never promised, is usually a
\emph{false alarm}, not a bug.

\textbf{PBT libraries.} PBT libraries provide building blocks for
generators and a runner that executes a test on many inputs; the user combines these building blocks into a generator and writes the oracle. Examples are
QuickCheck~\cite{claessen2000quickcheck} for Haskell and
Hypothesis~\cite{maciver2019hypothesis} for Python. In Hypothesis,
a generator is called a \emph{strategy}. The library provides strategies for
common types, such as \CodeIn{st.integers()} and \CodeIn{st.lists()}, which can
be combined into strategies for structured inputs. The \CodeIn{@given} decorator
turns a test function with an assertion into a property-based test that runs
the function on inputs drawn from the strategies. The \CodeIn{@settings}
decorator configures the run; for example, \CodeIn{@settings(max\_examples=N)}
sets how many inputs to draw, 100 by default. When the test fails, Hypothesis \emph{shrinks} the failing input to a smaller
one that still fails.

\begin{wrapfigure}{r}{0.42\linewidth}
\vspace{-1.2em}
\begin{tcblisting}{codebox,
  minted language=python,
  minted options={fontsize=\scriptsize}}
@given(st.lists(st.integers()))
def test_output_is_ordered(xs):
    ys = quicksort(xs)
    assert all(ys[i] <= ys[i + 1]
               for i in range(len(ys) - 1))
\end{tcblisting}
\vspace{-0.5em}
\caption{A Hypothesis test that checks that \CodeIn{quicksort}'s output is
ordered.}
\label{fig:bg-quicksort-ordered}
\vspace{-1em}
\end{wrapfigure}
\sloppy
\textbf{Example.} Figure~\ref{fig:bg-quicksort-ordered} shows a PBT in
Hypothesis for a function \CodeIn{quicksort} that sorts a list of integers. The
decorator \CodeIn{@given} tells Hypothesis to draw inputs from the
strategy \CodeIn{st.lists(st.integers())}, which generates lists of integers,
including the empty list and lists with duplicate or negative elements. The test calls \CodeIn{quicksort} on each
drawn list $xs$, and the oracle checks that the output
$ys$ is ordered, a property of every correct sorting implementation. If
\CodeIn{quicksort} misplaces an element, the assertion fails, and Hypothesis
reports a shrunk failing list as the counterexample. This oracle is partial.
A \CodeIn{quicksort} that drops elements still passes it; catching that bug
also requires checking that $ys$ is a permutation of $xs$.

\subsection{LLM-Based PBT Writers}

\label{sec:bg-llm}

\proptestai~\cite{vikram2024pbt} and \agenticpbt~\cite{maaz2025agentic} are two recent LLM-based PBT writers for Python. Figure~\ref{fig:bg-prompts}
shows their prompts.

\textbf{\proptestai.} \proptestai~\cite{vikram2024pbt} prompts an LLM with the
documentation of one Python API method and asks for PBTs of that method. Its
authors propose two prompting modes. The \emph{one-prompt} mode
uses one prompt, which asks for a list of properties and a test that checks
them in the same answer. The \emph{two-prompt} mode
(Figure~\ref{fig:bg-prompt-proptest}) splits this work into two prompts, sent
in sequence in one conversation. The Stage~1 prompt gives the method's
name (\promptparam{function\_name}) and its documentation from the
library's website (\promptparam{api\_documentation}), and asks the LLM
to list up to five properties of the method's output. The Stage~2 prompt,
sent after the LLM answers, asks for a Hypothesis test suite that checks the
properties the LLM just listed. It ends with a template that imports the
method's module (\promptparam{module\_name}) and names the test
function. The authors evaluate \proptestai on 40 deterministic API methods from ten Python
libraries, such as \CodeIn{numpy}, \CodeIn{pandas}, \CodeIn{networkx}, and
\CodeIn{datetime}, with GPT-4, Claude-3-Opus, and Gemini-1.5-Pro. In their best
configuration, GPT-4 with two prompts, 41.74\% of the tests from Stage~2 run
without error and pass on the library, against 31.04\% for the next best,
Claude-3-Opus with two prompts.

\newsavebox{\bgpromptleft}
\begin{figure}[t]
\centering
\begin{minipage}[t]{0.49\linewidth}
\vspace{0pt}
\begin{lrbox}{\bgpromptleft}
\begin{minipage}{\linewidth}
\begin{tcblisting}{partbox, title={Stage 1 prompt},
  minted language=text, minted options={fontsize=\tiny,breaklines,escapeinside=!!}}
You are a world class Python programmer.
Please review the following API documentation
for !\promptslot{\{function\_name\}}!.
Documentation: !\promptslot{\{api\_documentation\}}!
Extract a list of maximum five properties that
you would use for property-based testing.
Specifically, give properties on the output of
the function, which may or may not depend on
the input arguments. [...]
Do not write any tests. Only write this list
of five properties in natural language.
\end{tcblisting}
\vspace{4pt}
\begin{tcblisting}{partbox, title={Stage 2 prompt, sent after the LLM's reply},
  minted language=text, minted options={fontsize=\tiny,breaklines,escapeinside=!!}}
Now, write a Python property-based test suite
using the library Hypothesis testing the
properties you described. [...]
In your generation strategy, be aware of very
large inputs and overflows.
-------------- Output Format --------------
from hypothesis import given, strategies as st
import !\promptslot{\{module\_name\}}!

@given(st.data())
def test_!\promptslot{\{function\_name\}}!_property():
    (...)
# End program
\end{tcblisting}
\end{minipage}
\end{lrbox}
\xdef\bgpromptheight{\the\dimexpr\ht\bgpromptleft+\dp\bgpromptleft\relax}%
\usebox{\bgpromptleft}
\subcaption{\proptestai (two-prompt mode).}
\label{fig:bg-prompt-proptest}
\end{minipage}\hfill
\begin{minipage}[t]{0.49\linewidth}
\vspace{0pt}
\begin{tcblisting}{codebox,
  height=\bgpromptheight,
  minted language=text, minted options={fontsize=\tiny,breaklines,escapeinside=!!}}
# Property-Based Testing Bug Hunter
You are a bug-hunting agent focused on finding
genuine bugs through property-based testing
with Hypothesis. [...]
## Your Todo List
1. Analyze target  2. Understand the target
3. Propose properties  4. Write tests
5. Test execution and bug triage
6. Report or conclude
## Core Process
### 2. Understand the target [...]
### 3. Propose properties
- Invariants  - Round-trip properties
- Inverse operations  - Multiple implementations
- Metamorphic properties [...]
Only test properties that the code is
explicitly claiming to have. [...]
### 4. Write tests
- sound: tests only inputs expected by the code
- complete: tests all inputs expected by the code
If soundness and completeness are in conflict,
prefer writing sound but incomplete properties.
### 5. Test execution and bug triage
Run your tests with `pytest`.
For test failures, apply this bug triage rubric:
[...] - Do callers of this code make assumptions
  that prevent this input? [...]
If false alarm detected: Return to Step 4 [...]
### 6. Bug Reporting [report template] [...]
Now analyze the targets: !\promptslot{\$ARGUMENTS}!
\end{tcblisting}
\subcaption{\agenticpbt.}
\label{fig:bg-prompt-agentic}
\end{minipage}
\caption{The prompts of \proptestai~\cite{vikram2024pbt}, from its
artifact~\cite{vikram2024proptestdata}, and \agenticpbt~\cite{maaz2025agentic},
verbatim but abridged. \texttt{[...]}
marks cuts; shading marks the parts filled in per target.}
\label{fig:bg-prompts}
\end{figure}

\textbf{\agenticpbt.} \agenticpbt~\cite{maaz2025agentic} is a natural-language
prompt (Figure~\ref{fig:bg-prompt-agentic}) that directs a \emph{coding agent},
an LLM that runs commands and edits files, to write
property-based tests for Python code, run them, and report bugs. The
prompt takes one argument (\promptraw{\$ARGUMENTS}), the target: a Python
module, file, or function, such as \CodeIn{numpy.linalg}. It lists six steps.
The agent reads the target's source code and documentation (steps 1--2). It
proposes properties that the code claims in its docstrings, comments, or
callers, and reads the callers to learn which inputs the target expects (step
3). It writes a Hypothesis test for each property (step 4). A strategy should
draw only inputs the code expects; when it cannot also draw all of them, the
prompt prefers missing valid inputs to drawing invalid ones. The agent then runs
the tests with \CodeIn{pytest} and triages each failure with a rubric in the
prompt. The rubric asks whether the failure reproduces, whether callers can pass
the failing input, and whether the code claims the property (step 5). When the
agent judges a failure a false alarm, it revises the test, its strategy or its
property, and reruns it. It reports each remaining failure with the failing
test, the shrunk input, a reproducing script, and, when possible, a patch (step
6). The authors run \agenticpbt with Claude Opus~4.1 on 933 modules of 100 Python packages, from the standard library to
\CodeIn{numpy} and \CodeIn{pandas}. The agent reports 984 failures as likely bugs. The authors review 50 of these alarms, sampled from the 80\% that an LLM
ranks highest, and judge 28 to be real bugs.

\textbf{Differences.} Table~\ref{tab:bg-tools} compares the two tools. The
key difference is that only \agenticpbt runs its tests and revises those behind
false alarms; \proptestai leaves both to the user. They also differ in where they find
properties: \proptestai receives the documentation in its prompt, whereas
\agenticpbt searches the code, docstrings, comments, and callers itself.

\begin{table}[t]
\centering
\footnotesize
\caption{\proptestai and \agenticpbt as reported by their authors, whose results are not comparable.}
\label{tab:bg-tools}
\scriptsize\renewcommand{\arraystretch}{1.05}
\setlength{\aboverulesep}{0.2ex}\setlength{\belowrulesep}{0.3ex}
\begin{tabularx}{\linewidth}{>{\raggedright\arraybackslash}p{2.2cm}>{\raggedright\arraybackslash}X>{\raggedright\arraybackslash}X}
\toprule
 & \textbf{\proptestai}~\cite{vikram2024pbt} & \textbf{\agenticpbt}~\cite{maaz2025agentic} \\
\midrule
\rowcolor{black!8}
Target & One Python API method & A Python module, file, or function \\
Property source & Documentation, given in the prompt & Code, docstrings, comments, and callers it finds \\
\rowcolor{black!8}
LLM use & One prompt, or two in sequence & Agent loop with shell and file tools \\
Runs its tests & No & Yes, with \textit{pytest} \\
\rowcolor{black!8}
Revises its tests & No & Yes, when a triage rubric flags a false alarm \\
Output & Hypothesis test suite & Hypothesis test suite, plus a written report for each failure it judges a bug \\
\rowcolor{black!8}
Evaluation & 40 methods, 10 libraries;\newline GPT-4, Claude-3-Opus, Gemini-1.5-Pro & 933 modules, 100 packages;\newline Claude Opus 4.1 \\
Reported result & 41.74\% of tests run and pass\newline (best configuration) & 28 of 50 sampled alarms\newline judged real by the authors \\
\bottomrule
\end{tabularx}
\end{table}

%% file: domains.tex
\section{Testing the Domains with PBTs}
\label{sec:domains}

\begin{table}[!tp]
\centering
\scriptsize
\setlength{\tabcolsep}{4pt}
\renewcommand{\arraystretch}{1.05}
\setlength{\aboverulesep}{0.2ex}\setlength{\belowrulesep}{0.3ex}
\caption{How a test reaches the programs of each domain.}
\label{tab:domains}
\begin{tabularx}{\linewidth}{@{}>{\raggedright\arraybackslash}p{2.4cm}>{\raggedright\arraybackslash}p{1.0cm}>{\raggedright\arraybackslash}p{4.9cm}>{\raggedright\arraybackslash}X@{}}
\toprule
\textbf{Domain} & \textbf{Level} & \textbf{Input language} & \textbf{Harness} \\
\midrule
\textit{Prior evaluations} & API & Python values & None \\
\midrule
\rowcolor{black!8}
Deep learning libraries & API & Python values (tensors) & None \\
Databases & System & SQL; AQL & Starts a server, sends queries \\
\rowcolor{black!8}
System software & System & Configuration text & Runs a command; starts sshd \\
Compilers & System & C and C++ programs & Compiles and runs a program \\
\rowcolor{black!8}
LLM serving & API & \drafttext{Python values, prompts, and configurations} & None \\
\bottomrule
\end{tabularx}
\end{table}

A domain is a class of programs that take the same kind of input and that a
test controls and observes the same way~\cite{Amman_Offutt_2016}, such as
through a function call or a process. This section describes how a PBT reaches and checks the programs of each of
the \numDomains{} domains we study, and shows example tests that \agenticpbt wrote.

\subsection{Test Characteristics}
\label{sec:domains-chars}

Table~\ref{tab:domains} describes the tests of each domain by three
characteristics, which the domain and its focal points fix. \emph{Level} is where a test meets the program: an API-level test
calls a function, and a system-level test runs a program or sends input to a
running one. \emph{Input language} is the form of the test's input, from
Python values to text in the program's own language, such as SQL, C, or a
configuration file. \emph{Harness} is the code, beyond the generator and oracle, that builds,
starts, or feeds the program under test. API-level tests call the program
directly and have no harness.

What separates our domains from the prior evaluations are the three characteristics of Table~\ref{tab:domains}. Deep learning libraries match the prior
evaluations in all three. LLM serving also uses API calls without a harness,
with inputs ranging from Python values to prompts and configurations.
Databases, system software, and compilers differ in all three. In these three domains, the harness is one more function to call, since we
write it and the writer reads its code, but the input is text in the
program's own language, valid only under the program's semantics, and the
oracle must know what that text means.  The prior evaluations cover up to 933 Python modules~\cite{maaz2025agentic},
all called with Python values; our ten programs are fewer, but a sharper test
of whether a writer generalizes.

The tests of Section~\ref{sec:domains-studied} rely on two further notions. A
constraint on the inputs is \emph{relational} when it relates one input to
another. Because Hypothesis's built-in strategies draw each value on its own,
a relational constraint usually requires a \emph{custom generator}: built-in
strategies combined with \CodeIn{filter} or \CodeIn{map}, or a function
written with \CodeIn{@st.composite}. We name oracle families after the survey of Barr et
al.~\cite{barr2015oracle}, whose ``assertions and contracts'' and
``pseudo-oracles'' are our first two. A
\emph{contract} oracle checks one call against a condition on its outcome,
including an expected output that the test computes from the input. A
\emph{differential} oracle compares the output with that of an independent
implementation; it can be another program, another function or a different mode of the
same program. A \emph{metamorphic} oracle relates the outputs of two calls whose inputs differ in a known way. RQ2 (Section~\ref{sec:rq2}) adds two rarer families from the same survey.

\subsection{Studied Domains}
\label{sec:domains-studied}

We select the \numDomains{} domains below for their popularity and so that
they vary in Level, Input language, and Harness.  For each of the three domains that
differ from the prior evaluations, a figure shows one test that \agenticpbt
wrote, because these tests show what the writer must produce beyond a Python
call: a query, a crontab, or a C++ program, and an oracle for it. Each figure
omits the test's imports, helper code, assertion messages, and checks that the
program neither crashed nor timed out, and renames its generator
\CodeIn{custom\_gen}. A generator marked \emph{simplified} omits values that the test never uses
(a partition column in PostgreSQL's rows, a flag in Clang's tuple) and some
forms of input: frames without an \texttt{EXCLUDE} clause in PostgreSQL, and
month and weekday names (e.g., \texttt{jan}) and lists with an asterisk
(e.g., \texttt{*,5}) in cronie.
For the other two domains, whose tests resemble those of the prior
evaluations, we describe a test in prose.
Each of the three tests shown in a figure checks a constraint stated in the
program's documentation, needs a custom generator to meet it, and fails on
the program.

\label{sec:domains-dl}

\par\textbf{Deep Learning Libraries.}~JAX~\cite{jax2018github} and PyTorch~\cite{paszke2019pytorch} are Python
libraries for building and training neural networks, and many techniques test
such
libraries~\cite{Zhang_ETAL_TSE22,Phan_ETAL_ICSE2019,Wei2022,deng2022fuzzing,
Xie_ETAL_ISSTA22,deng2023titanfuzz,Liu_2023}.
Their APIs take as inputs \emph{tensors}, which are multi-dimensional arrays
with a shape and an element type. A test calls an API directly. For example, \agenticpbt's test of
\texttt{jax.lax.pmean}, which averages a tensor over groups of devices, must
draw groups that ``cover all axis indices exactly once''~\cite{jaxpmean},
which a randomly drawn list of lists rarely does. Its custom generator draws
one group label per device and groups the devices with the same label, and
its oracle computes the mean of each group in NumPy and compares it with the
output of the API.

\label{sec:domains-db}
\begin{figure}[!tb]
\centering
\begin{tcblisting}{codebox,
  minted language=python, minted options={fontsize=\scriptsize, escapeinside=||}}
# Property-based test
@settings(max_examples=400, deadline=None)
@given(rows=tables, frame=|\codehl{custom\_gen}|(), order=orderings, n=st.integers(1, 6))
def test_nth_value_matches_frame(rows, frame, order, n):
    r = run(f"SELECT nth_value(v, {n}) OVER w, array_agg(v) OVER w "
            f"FROM ({values_clause(rows)}) t(k, v) WINDOW w AS (ORDER BY {order} {frame})")
    assume(r.ok)
    for nv, agg in r.rows:
        reference = agg[n - 1] if len(agg or []) >= n else None
        assert nv == reference
# Custom generator (simplified)
@st.composite
def custom_gen(draw):
    kinds = ["UNBOUNDED PRECEDING", "{} PRECEDING", "CURRENT ROW", "{} FOLLOWING", "UNBOUNDED FOLLOWING"]
    start = draw(st.integers(0, 3))
    end = draw(st.integers(1, 4).filter(lambda e: e >= start))
    mode = draw(st.sampled_from(["ROWS", "RANGE", "GROUPS"]))
    excl = "EXCLUDE " + draw(st.sampled_from(["NO OTHERS", "CURRENT ROW", "GROUP", "TIES"]))
    lo, hi = kinds[start].format(draw(st.integers(0, 4))), kinds[end].format(draw(st.integers(0, 4)))
    return f"{mode} BETWEEN {lo} AND {hi} {excl}"
keys, vals = st.one_of(st.integers(0, 4), st.none()), st.one_of(st.integers(-9, 9), st.none())
tables = st.lists(st.tuples(keys, vals), min_size=1, max_size=9)
orderings = st.sampled_from(["k", "k DESC", "k NULLS FIRST", "k DESC NULLS LAST"])
# Example input (reveals a failure)
rows = [(0, None), (0, 0), (None, 0)]
frame = "ROWS BETWEEN 1 FOLLOWING AND UNBOUNDED FOLLOWING EXCLUDE TIES"
order, n = "k", 1
\end{tcblisting}
\caption{\textbf{PostgreSQL (database).} \agenticpbt's test for \texttt{nth\_value}. \texttt{run} sends a query to
the server through the harness, and \texttt{values\_clause} writes the rows as a \texttt{VALUES}
list. Each row holds a sort key \texttt{k} and a value \texttt{v}.}
\label{fig:dom-db}
\end{figure}

\par\textbf{Databases.}~A database server answers queries over its tables, and testing database
servers has its own line of
research~\cite{slutz1998rags,rigger2020pqs,rigger2020norec,rigger2020tlp,zhong2020squirrel}.
We study PostgreSQL~\cite{postgresql}, queried in SQL, and
ArangoDB~\cite{arangodb}, queried in AQL; a test writes a query as text, sends it to a running server, and checks the
rows. Figure~\ref{fig:dom-db} shows a test of
PostgreSQL's \texttt{nth\_value(v, n)}, which returns the value in the $n$-th
row of a \emph{window frame}, a range of rows around each row. The test fills a query template with a table, a sort order, a frame clause,
and a row index $n$.
Built-in strategies draw all but the frame clause, which needs a custom
generator, \CodeIn{custom\_gen}, because its two ends constrain each other.
The generator orders the kinds of the two ends, as the manual requires when it
states that ``the \texttt{frame\_end} choice cannot appear earlier [\ldots]
than the \texttt{frame\_start} choice does''~\cite{postgresqlwindow}. Sort keys draw from five values and \texttt{NULL}, so rows often tie, as in
the example input (that reveals a failure); ties matter, in particular because the frame's \texttt{EXCLUDE} options drop
the current row or the rows tied with it.
The test discards queries the server rejects, so it cannot detect a server
that wrongly rejects a valid query. Its oracle is differential: it computes \texttt{array\_agg(v)}, another function
of the server, over the same frame in the
same query and checks that \texttt{nth\_value}
equals its $n$-th element, or \texttt{NULL} when the frame holds fewer than
$n$ rows. The test fails for $n = 1$ and the frame \texttt{ROWS BETWEEN 1
FOLLOWING AND UNBOUNDED FOLLOWING EXCLUDE TIES}: for the first of two tied rows
that a third row follows, \texttt{array\_agg} returns the third row's value,
but \texttt{nth\_value} returns \texttt{NULL}~\cite{postgresql-19732}.

\label{sec:domains-sys}
\begin{figure}[!tb]
\centering
\begin{tcblisting}{codebox,
  minted language=python, minted options={fontsize=\scriptsize, escapeinside=||}}
# Property-based test
@settings(max_examples=300, deadline=None)
@given(|\codehl{custom\_gen}|, st.integers(1_000_000_000, 2_000_000_000), st.integers(60, 5 * 24 * 3600))
def test_all_jobs_in_interval_match_documented_semantics(fields, start, window):
    r = cmd_harness.run("cronnext", ["-t", str(start), "-q", str(start + window), "-f", "tab"],
                        files={"tab": " ".join(fields) + " echo job\n"})
    reference = Entry(*fields).matches_in((start + 59) // 60 * 60, (start + window) // 60 * 60)
    assert sorted(int(m) for m in NEXT_RE.findall(r.stdout)) == reference
# Custom generator (simplified)
def field(lo, hi):
    rng = st.tuples(st.integers(lo, hi), st.integers(lo, hi)).filter(lambda ab: ab[0] <= ab[1])
    rng = rng.map(lambda ab: "%
    stepped = st.tuples(rng, st.integers(2, hi - lo)).map(lambda rs: "%
    plain = st.one_of(st.integers(lo, hi).map(str), rng, stepped)
    star = st.one_of(st.just("*"), st.integers(2, hi - lo).map(lambda s: "*/%
    return st.one_of(star, st.lists(plain, min_size=1, max_size=3).map(",".join))
custom_gen = st.tuples(field(0, 59), field(0, 23), field(1, 31), field(1, 12), field(0, 7))
# Example input (reveals a failure)
fields = ("*", "*", "*", "*", "*/2")
start, window = 1_000_000_000, 80000
\end{tcblisting}
\caption{\textbf{cronie (system software).} \agenticpbt's test for the focal point ``cron command lines,'' a section of
crontab(5). \texttt{Entry} is the model of crontab(5), and
its method \texttt{matches\_in} lists the minutes at which the job runs;
\texttt{NEXT\_RE} matches a time in the output of \texttt{cronnext}, and
\texttt{cmd\_harness} is the harness.}
\label{fig:dom-sys}
\end{figure}

\par\textbf{System Software.}~OpenSSH~\cite{openssh} and cronie~\cite{cronie} are Unix programs for remote
login and job scheduling, and system software of this kind has been a target of
fuzzing research~\cite{pham2020aflnet}. Each program reads a configuration
file, such as \texttt{sshd\_config} or a crontab, in a format that its manual
page defines. A test writes a configuration file, runs a program on it, and
parses what the program prints.
Figure~\ref{fig:dom-sys} shows a test of cronie's program \texttt{cronnext},
which lists the times in a window at which a crontab's job runs. The test
fills a crontab line template with five generated time fields and draws the window with built-in strategies. A field can be an asterisk, a number, a range,
a step, or a list of these, and crontab(5) states that in a range ``the first
number must be less than or equal to the second one''~\cite{croniecrontab}.
The test therefore uses a custom generator, \CodeIn{custom\_gen}, that draws each field within its bounds, keeps only a range whose ends are in
order, and draws a step no larger than the field's span; the figure ends with the draw on which the test fails. The test writes
the line to a file and runs \texttt{cronnext} on it through the harness. Its
oracle is a contract, because code of the test computes the expected output:
the test compares the times that \texttt{cronnext} lists with
those that a 92-line Python model of crontab(5) computes, which \agenticpbt{}
also wrote. The test fails on the line \texttt{* * * * */2}. crontab(5) states
that ``Step values are also permitted after an asterisk,'' so \texttt{*/2} in
the day-of-week field names every second day from Sunday: Sunday, Tuesday,
Thursday, and Saturday. The job must therefore not run on a Monday, but for a
window that starts on a Sunday, \texttt{cronnext} also lists midnight of the
following Monday~\cite{cronie-215}.

\label{sec:domains-cc}
\begin{figure}[!tb]
\centering
\begin{tcblisting}{codebox,
  minted language=python, minted options={fontsize=\scriptsize, escapeinside=||}}
# Property-based test
@settings(max_examples=100, deadline=None)
@given(|\codehl{custom\_gen}|())
def test_ptr_align_constfold_matches_runtime(c):
    ba, al, sz, off, op = c
    reference = (off + al - 1) // al * al if op == "up" else off // al * al
    assume(0 <= reference < sz)
    r = cc_harness.compile_and_run(PROG %
    assume(r.compiled)
    ct_diff, ct_read, rt_diff, rt_read = map(int, r.run_stdout.split())
    assert (rt_diff, rt_read) == (reference, reference %
    assert (ct_diff, ct_read) == (rt_diff, rt_read)
# Custom generator (simplified)
@st.composite
def custom_gen(draw):
    ba = 1 << draw(st.integers(0, 7))
    al = 1 << draw(st.integers(0, 7))
    sz = draw(st.integers(1, 4)) * ba
    off = draw(st.integers(0, sz - 1))
    return ba, al, sz, off, draw(st.sampled_from(["up", "down"]))
# Example input (reveals a failure)
ba, al, sz, off, op = 2, 2, 2, 1, "down"
\end{tcblisting}
\begin{tcblisting}{codebox,
  minted language=cpp, minted options={fontsize=\scriptsize}}
// Failing input, rendered: the program that PROG yields for the values above
struct S { alignas(2) unsigned char d[2]; };
constexpr S g = mk();                                    // g.d[i] == i
constexpr const unsigned char *cp = __builtin_align_down(g.d + 1, 2);
constexpr int ct_diff = (int)(cp - g.d);
constexpr int ct_read = (int)*cp;
__attribute__((noinline)) const unsigned char *rt(const unsigned char *p, unsigned long a) {
  return __builtin_align_down(p, a);
}
volatile int voff = 1;
int main() {
  const unsigned char *p = rt(g.d + voff, 2);
  printf("%
}
\end{tcblisting}
\caption{\textbf{Clang (compiler).} \agenticpbt's test for ``Alignment builtins''
in Clang's manual, and below, the program that the template \texttt{PROG}
yields for the failing input (\texttt{\#include} and \texttt{mk}'s body
omitted).}
\label{fig:dom-cc}
\vspace{-0.8em}
\end{figure}

\par\textbf{Compilers.}~Clang~\cite{clang} and GCC~\cite{gcc} compile C and C++ programs, and many
techniques generate programs to test
compilers~\cite{yang2011csmith,le2014emi,livinskii2020yarpgen,chen2020compilersurvey}.
A test compiles and runs a small program, which must be free of
\emph{undefined behavior}, because the C and C++ standards allow any output
for such a program. Figure~\ref{fig:dom-cc} shows a test of Clang's
\texttt{\_\_builtin\_align\_up} and \texttt{\_\_builtin\_align\_down}, which
round a pointer to a multiple of an alignment and, the manual states, ``are
evaluatable inside constant expressions''~\cite{clangalign}. The test fills a C++ program template, \texttt{PROG}, with five generated values: the
alignment and size of an array whose byte $i$ holds $i$ modulo 256, an offset
into it, the alignment to round to, and the direction. The program rounds the
pointer at compile time and at run time and prints, for each, the result's
offset and the byte it points to. The manual states that if the alignment ``is not a power of two at run time,
the behavior of these builtins is undefined,'' and that the result ``must be
within the same underlying allocation or one past the end''~\cite{clangalign}.
The custom generator, \CodeIn{custom\_gen}, therefore draws both alignments as powers of two, which \texttt{alignas} also
requires of the array's, the size as a multiple of the array's alignment, and
an offset inside the array. The test discards a draw whose result leaves the array, even one past the
end, because the program reads through the pointer. It also discards a program
that fails to compile, so it cannot detect a compiler that wrongly rejects a
valid program. Its oracle is differential: the test requires the compile-time results to
equal the run-time ones. It also computes the rounded offset in Python and
requires the run-time offset and byte to match it, which is a contract. The
test fails when \texttt{ba}, \texttt{al}, and \texttt{sz} are 2, \texttt{off}
is 1, and \texttt{op} is \texttt{down}: both offsets are 0 and the run-time
read returns byte 0, but the compile-time read returns byte 1~\cite{llvm-227548}.

\label{sec:domains-llm}

\par\textbf{LLM Serving.}~\drafttext{An LLM serving engine answers prompts with a
model: it batches requests, applies the fine-tuned weights that a request asks
for, and streams each response as it is generated.} Testing serving engines is
a recent line of work~\cite{liu2026firstlook,zhao2026grief}. \drafttext{We
study vLLM~\cite{kwon2023vllm} and SGLang~\cite{zheng2024sglang}, whose tests
call Python APIs of the engines and their components or start a whole engine.
For example, \agenticpbt checks that a response does not depend on the other
requests that the engine handles~\cite{vllm-57493}, that streaming
does not change it~\cite{sglang-35563}, and that the engine applies the
fine-tuned weights that it accepts~\cite{vllm-55193}. Because a response has no
single correct value, these properties compare two runs whose responses must
be equal, as with different batches or streaming modes, or must differ, as
with and without the fine-tuned weights.}

%% file: eval.tex
\section{Evaluation}
\label{sec:eval}

\newcommand{\rqone}{How do the two LLM-based PBT writers compare in coverage and failures?}
\newcommand{\rqtwo}{What characterizes the property-based tests that \agenticpbt writes?}
\newcommand{\rqtwoa}{Which families of oracle do the tests check the program with?}
\newcommand{\rqtwob}{Which artifacts state the properties that the tests check?}
\newcommand{\rqthree}{How useful is \agenticpbt in finding developer-confirmed bugs?} %
\newcommand{\rqfour}{How much do \agenticpbt's tests and domain-specific fuzzers cover, and which crashes does each find that the other does not?}

We evaluate LLM-based PBT writers with the following research questions:
\begin{packed_itemize}
\item \textbf{RQ1} [\textbf{Performance}]: \textit{\rqone}
\item \textbf{RQ2} [\textbf{Characterization}]: \textit{\rqtwo}
\item \textbf{RQ3} [\textbf{Usefulness}]: \textit{\rqthree}
\item \textbf{RQ4} [\textbf{Domain-Specific Tools}]: \textit{\rqfour}
\end{packed_itemize}

RQ1 evaluates the \emph{performance} of the writers across the five domains
and selects one to study further. RQ2 \emph{characterizes} the tests by the families of oracles they check and
the artifacts that state their properties. RQ3 measures \emph{usefulness} by the
bugs that developers confirm. RQ4 compares \agenticpbt with domain-specific tools in branch coverage and
in the crashes that each reveals.

\emph{Targets and focal points.} Table~\ref{tab:subjects} lists the ten
programs of Section~\ref{sec:domains-studied}, two per domain, which we call
\emph{targets}, and the version of each that we test; RQ3 tests older versions of the same
targets (Table~\ref{tab:bugs}). A writer tests
a target one focal point at a time. Its kind depends on the domain: an API
method or class of a Python library, a function of a query language, an
option or keyword of a configuration file, a construct of the crontab syntax,
or a \drafttext{language extension} of a compiler. \drafttext{We extract each target's focal
points from the documentation in the table's Source column}, and the last
column gives their number.

\begin{table}[t]
\centering
\scriptsize
\renewcommand{\arraystretch}{0.95}\setlength{\aboverulesep}{0.2ex}\setlength{\belowrulesep}{0.3ex}
\caption{The ten targets of RQ1 and RQ2 (RQ4 uses four of them), \drafttext{the documentation from which
we extract their focal points}, one example, and \drafttext{their number}.}
\label{tab:subjects}
\begin{tabular}{@{}lllllr@{}}
\toprule
& & & \multicolumn{3}{c}{\textbf{Focal point}} \\
\cmidrule(l){4-6}
\textbf{Domain} & \textbf{Target} & \textbf{Version} & \textbf{\drafttext{Source}} & \textbf{\drafttext{Example}} & \textbf{Number} \\
\midrule
\multirow{2}{*}{Deep learning} & JAX & 0.11.2 & \drafttext{\href{https://docs.jax.dev/en/latest/jax.html}{API reference}} & \texttt{jax.numpy.isin} & \draftnum{1,558} \\
                               & PyTorch & 2.14.0 & \drafttext{\href{https://docs.pytorch.org/docs/2.14/pytorch-api.html}{API reference}} & \texttt{torch.cumsum} & \draftnum{4,107} \\
\midrule
\multirow{2}{*}{LLM serving} & vLLM & 0.30.0 & \drafttext{\href{https://docs.vllm.ai/en/v0.30.0/api/}{API reference}} & \texttt{RepetitionDetectionParams} & \draftnum{4,399} \\
                             & SGLang & 0.5.20 & \drafttext{\href{https://github.com/sgl-project/sglang/tree/v0.5.20/python/sglang}{docstrings}} & \texttt{Siglip2Encoder} & \draftnum{5,716} \\
\midrule
\multirow{2}{*}{Database} & PostgreSQL & 18.6 & \drafttext{\href{https://www.postgresql.org/docs/18/functions.html}{function reference}} & \texttt{jsonb\_each\_text} & 634 \\
                          & ArangoDB & 3.12.12 & \drafttext{\href{https://docs.arangodb.com/3.12/aql/functions/}{function reference}} & \texttt{DATE\_ISOWEEK} & 255 \\
\midrule
\multirow{2}{*}{System software} & OpenSSH & 10.5p1 & \drafttext{\href{https://www.openssh.com/manual.html}{manual pages}} & \texttt{AddressFamily} & 167 \\
                                 & cronie & 1.7.2 & \drafttext{\href{https://man7.org/linux/man-pages/man5/crontab.5.html}{manual page}} & \texttt{@yearly} & 39 \\
\midrule
\multirow{2}{*}{Compiler} & GCC & 17 (\textit{dc301f9}) & \drafttext{\href{https://gcc.gnu.org/onlinedocs/gcc/}{manual}} & nested functions & \draftnum{1,000} \\
                          & Clang & 24 (\textit{4057f8e}) & \drafttext{\href{https://clang.llvm.org/docs/index.html}{manual}} & \texttt{fallthrough} & \draftnum{487} \\
\bottomrule
\end{tabular}
\end{table}

\subsection{RQ1: Performance of PBT Writers}
\label{sec:rq1}
\label{sec:rq1a} %

RQ1 compares \proptestai and \agenticpbt on ten programs in five domains by
their tests, cost, branch coverage, and failures, and selects the writer that
RQ2 to RQ4 study.

\newcommand{\numfocal}{30}
\newcommand{\numfocalpt}{50}  %
\newcommand{\fuzzbudget}{30 minutes on 20 CPU cores}

\par
\textbf{Setup.}
\label{sec:rq1-setup}
\emph{Sampling focal points.}
For each target, we sample \numfocal\ of its documented focal points
(Table~\ref{tab:subjects}) uniformly at random; on PyTorch, the 30 are the first 30 of the 50 APIs
that RQ4 draws uniformly from the 368 that both of its fuzzers support. Both
writers test the same focal points, which the artifact lists.
\emph{Writers.} We run both writers with two LLMs: Claude Opus~5, a proprietary frontier model, and GLM-5.3-Flash~\cite{glm53flash},
an open-weights model from another vendor. We chose GLM-5.3-Flash because it comes from a vendor other than Anthropic,
whose researchers developed \agenticpbt, because its weights are open, so
others can rerun the study without a proprietary API, and because it costs
much less to run. The second model tests whether the comparison between the writers holds
with another model. It does not test whether the writers perform equally
well with both.
\proptestai receives the documentation of each focal point. Following its
original evaluation~\cite{vikram2024pbt}, we use its two-prompt mode
(Section~\ref{sec:bg-llm}) and repeat the procedure five times per focal
point.
\par
\begin{wraptable}{r}{0.52\linewidth}
\vspace{-0.2em}
\centering
\scriptsize
\caption{Valid PBTs and LLM cost (\$) per writer and LLM.}
\label{tab:rq1a-cost}
\setlength{\tabcolsep}{1.6pt}\renewcommand{\arraystretch}{0.95}
\setlength{\aboverulesep}{0.2ex}\setlength{\belowrulesep}{0.3ex}
\begin{tabular}{lrrrrrrrr}
\toprule
& \multicolumn{4}{c}{\textbf{Opus~5}} & \multicolumn{4}{c}{\textbf{GLM-5.3-Flash}} \\
\cmidrule(lr){2-5}\cmidrule(lr){6-9}
& \multicolumn{2}{c}{\agenticpbt} & \multicolumn{2}{c}{\proptestai} & \multicolumn{2}{c}{\agenticpbt} & \multicolumn{2}{c}{\proptestai} \\
\cmidrule(lr){2-3}\cmidrule(lr){4-5}\cmidrule(lr){6-7}\cmidrule(lr){8-9}
\textbf{Target} & \textbf{Tests} & \textbf{\$} & \textbf{Tests} & \textbf{\$} & \textbf{Tests} & \textbf{\$} & \textbf{Tests} & \textbf{\$} \\
\midrule
JAX & \draftnum{315} & \draftnum{51.7} & \draftnum{677} & \draftnum{41.7} & \draftnum{225} & \draftnum{1.1} & \draftnum{549} & \draftnum{0.1} \\
PyTorch & \draftnum{436} & \draftnum{54.7} & \draftnum{661} & \draftnum{30.9} & \draftnum{241} & \draftnum{0.7} & \draftnum{610} & \draftnum{0.1} \\
vLLM & \draftnum{202} & \draftnum{62.5} & \draftnum{700} & \draftnum{45.1} & \draftnum{150} & \draftnum{1.0} & \draftnum{482} & \draftnum{0.1} \\
SGLang & \draftnum{191} & \draftnum{68.2} & \draftnum{540} & \draftnum{49.6} & \draftnum{173} & \draftnum{1.2} & \draftnum{364} & \draftnum{0.1} \\
PostgreSQL & \draftnum{256} & \draftnum{73.0} & \draftnum{733} & \draftnum{75.3} & \draftnum{173} & \draftnum{1.3} & \draftnum{612} & \draftnum{0.1} \\
ArangoDB & \draftnum{244} & \draftnum{59.5} & \draftnum{699} & \draftnum{51.6} & \draftnum{117} & \draftnum{1.4} & \draftnum{618} & \draftnum{0.1} \\
OpenSSH & \draftnum{204} & \draftnum{64.7} & \draftnum{702} & \draftnum{87.0} & \draftnum{170} & \draftnum{2.0} & \draftnum{617} & \draftnum{0.1} \\
cronie & \draftnum{142} & \draftnum{66.7} & \draftnum{716} & \draftnum{90.8} & \draftnum{135} & \draftnum{1.5} & \draftnum{561} & \draftnum{0.1} \\
GCC & \draftnum{218} & \draftnum{87.7} & \draftnum{722} & \draftnum{76.9} & \draftnum{135} & \draftnum{1.4} & \draftnum{509} & \draftnum{0.1} \\
Clang & \draftnum{144} & \draftnum{81.9} & \draftnum{703} & \draftnum{96.8} & \draftnum{114} & \draftnum{1.3} & \draftnum{557} & \draftnum{0.1} \\
\midrule
\textbf{Total} & \draftnum{2,352} & \draftnum{670.6} & \draftnum{6,853} & \draftnum{645.5} & \draftnum{1,633} & \draftnum{12.9} & \draftnum{5,479} & \draftnum{1.1} \\
\bottomrule
\end{tabular}
\vspace{0.2em}
\end{wraptable}
\emph{Harness.} On the six targets outside Python, both writers
use the harness of the target's domain (Table~\ref{tab:domains})\drafttext{; for
\proptestai, we append a description of the harness to the documentation,
without changing its prompts}. We wrote each harness in about 30 minutes with the help of an LLM. The three
harnesses total 511 lines of Python, 1.8\% of the test code that \agenticpbt
writes for these targets.
\emph{Execution environment.} Each \emph{session}, one run of the agent on
one focal point, runs in its own container. The agent can read the target's source code, build, and harness, and can
write only to its working directory. RQ1 thus compares the writers as shipped:
\agenticpbt holds strictly more information than \proptestai, since it reads
the code and the harness and runs its tests with execution feedback, whereas
\proptestai sees only the documentation and the harness description, in five
two-prompt runs.
\emph{Budget.} \drafttext{Each writer fuzzes its tests of a target for
\fuzzbudget, split evenly among its tests, so both writers spend the same
fuzzing time on each target. We rerun each PBT with a new seed after it
finishes or fails, until its allocated time ends.} Each generation session is capped at 15 minutes. We do not cap the LLM cost,
and we report it.

\begin{figure}[t]
\centering
\includegraphics[width=\linewidth]{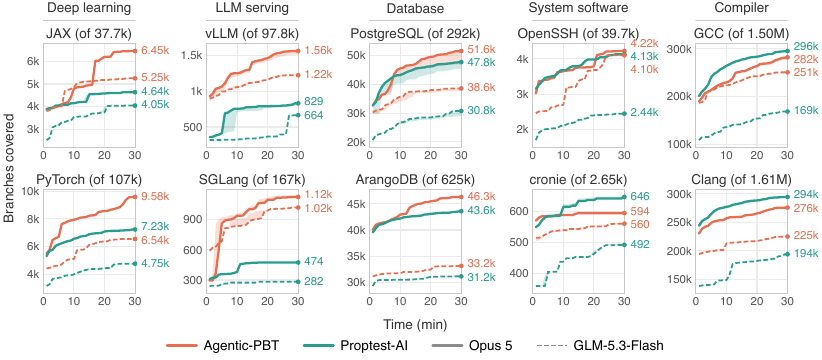}
\caption{Branch coverage over time, Opus~5 solid and GLM-5.3-Flash dashed; \drafttext{a band spans four fuzzing runs}. Titles give each target's total branches (Python code for JAX, vLLM, and SGLang; \texttt{libtorch\_cpu} for PyTorch).}
\label{fig:rq1a-cov}
\end{figure}

\par\textbf{Number of Tests Generated and Cost.}~Table~\ref{tab:rq1a-cost} gives, per target, the valid tests
that each writer produces\drafttext{ (property-based tests that pytest can
run)} and the LLM cost of producing them, with each LLM.
For \proptestai, we sum the valid PBT functions across its five responses
per focal point.
With Opus~5, \agenticpbt writes about \drafttext{a third} as many tests as \proptestai,
\draftnum{2,352} against \draftnum{6,853}, at a similar cost, \draftnum{\$670.6}
against \draftnum{\$645.5}, or about \draftnum{\$2.2} per focal point for each.
\drafttext{With GLM-5.3-Flash, \agenticpbt writes 1,633 tests for \$12.9 and
\proptestai 5,479 for \$1.1.} \drafttext{With Opus~5, \agenticpbt also writes
example-based tests for 41 of the 300 focal points, such as bug reproductions.}

\par\textbf{Coverage.}~We measure the branch coverage of each target as a whole,
on coverage-instrumented builds (coverage.py for Python code, gcov or
llvm-cov for C and C++). Figure~\ref{fig:rq1a-cov} plots the coverage of
each target over the fuzzing time, with the branches covered on the y-axis and
the minutes on the x-axis. It has one column per domain, with the domain's two
targets one above the other. The color of a line gives the writer, and its style gives the LLM,
solid for Claude Opus~5 and dashed for GLM-5.3-Flash. A
dot ends each line and is labeled with its final coverage. With Opus~5, \agenticpbt covers more branches than \proptestai on seven of
the ten targets: by \draftnum{33\%} to \draftnum{136\%} on the four API-level
targets, and by \draftnum{2\%} to \draftnum{8\%} on PostgreSQL, ArangoDB, and
OpenSSH, where the margin on OpenSSH is barely above the variation between fuzzing
runs reported below. \proptestai covers \draftnum{5\%} to \draftnum{9\%} more
on GCC, Clang, and cronie. On GCC, switching to GLM-5.3-Flash reduces
\proptestai's coverage from
\draftnum{296k} to \draftnum{169k} branches, against \draftnum{282k} to
\draftnum{251k} for \agenticpbt. With GLM-5.3-Flash, both writers
cover less, and \agenticpbt covers more than \proptestai on all ten targets.

Each curve is one generation run. Generating the tests of both writers with
Opus~5 costs about \$1,300 in LLM fees (Table~\ref{tab:rq1a-cost}), so we do
not repeat the generation. We instead repeat the fuzzing, which costs only
machine time, and report the margins between the writers. We use the second
LLM to examine how the comparison changes with model choice.
We rerun the fuzzing of every target's tests three times with new seeds, and the bands in Figure~\ref{fig:rq1a-cov} span the
four runs. These repetitions measure fuzzing variability for the generated
tests. With Opus~5, a writer's final coverage varies by at most
\draftnum{1.7\%} on eight targets, by \draftnum{4.8\%} on vLLM, and by \draftnum{10.1\%} on
PostgreSQL, where one run of \agenticpbt covers 9\% less than the other
three; the writers' ranges overlap only on PostgreSQL. \drafttext{With
GLM-5.3-Flash, a writer's coverage varies by at most \draftnum{1.2\%} on nine
targets and by \draftnum{6.6\%} on PostgreSQL, and no ranges overlap.}

\begin{wraptable}{r}{0.36\linewidth}
\vspace{-1.2em}
\centering
\footnotesize
\caption{Unique failures and sampled true-positive (TP) rates, with Opus~5;
30 failures sampled per writer.}
\label{tab:rq1-failures}
\setlength{\tabcolsep}{3pt}
\begin{tabular}{lrr}
\toprule
\textbf{Writer} & \textbf{Unique} & \textbf{TP (\%)} \\
\midrule
\agenticpbt & \draftnum{627} & \draftnum{30} \\
\proptestai & \draftnum{1,298} & \draftnum{13} \\
\bottomrule
\end{tabular}
\vspace{-1em}
\end{wraptable} 
\par\textbf{Failures.}~\drafttext{We examine failures from the initial fuzzing run with Opus~5,
which gives both writers higher coverage. Within each PBT, we deduplicate
failures by exception type and error message. We replace variable parts of
messages, such as numbers and memory addresses, with \texttt{\#}. For example,
\texttt{rows: [(1.0,), (0.0,)]} becomes
\texttt{rows: [(\#,), (\#,)]}.
For crashes, we compare signals and stack traces.
After setting the criteria, two authors independently
classify 30 randomly sampled unique failures per writer as true
or false positives.
Their labels show substantial agreement
($\kappa$ = \draftnum{0.69})~\cite{landis1977measurement}.
We resolve disagreements by discussing the evidence for each label.
We judge a failure to be a true positive when it violates documented behavior,
exposes a security defect, or disagrees with another implementation where both
specify the same behavior. We exclude incorrect test expectations, including
expectations about behavior left unspecified by the documentation. A true-positive
example is \CodeIn{torch.special.hermite\_polynomial\_he}, which violates its documented recurrence.
We use the final labels to calculate the true-positive rates.
Table~\ref{tab:rq1-failures} reports the unique failures and sampled
true-positive rates: 9 of 30 for \agenticpbt and 4 of 30 for \proptestai.
The 95\% Wilson confidence intervals are \draftnum{16.7--47.9\%} and
\draftnum{5.3--29.7\%}, respectively~\cite{wilson1927interval}.
Scaling the rates and their intervals by the unique failure counts gives
about \draftnum{190} (\draftnum{104--300}) and \draftnum{170}
(\draftnum{69--385}) true positives, respectively.}

\lessonblock{RQ1}{\agenticpbt covers more branches than \proptestai on all ten
targets with GLM-5.3-Flash and on seven with Opus~5, where \proptestai leads by \draftnum{5--9\%} on GCC, Clang, and
cronie. With Opus~5, \agenticpbt writes about a third as many tests.
Its sampled true-positive rate is \draftnum{30\%}, against \draftnum{13\%}
for \proptestai, but the small samples cannot confirm a difference.
We select \agenticpbt for its higher coverage on most targets and because
it alone runs and triages its tests, and Opus~5 for its higher coverage than
GLM-5.3-Flash.}

\FloatBarrier  %
\subsection{RQ2: Characteristics of the Property-Based Tests}
\label{sec:rq2}
\label{sec:rq2a} %
\label{sec:rq2b}

RQ2 characterizes the property-based tests that \agenticpbt writes with
Opus~5, the configuration that RQ1 selects. We measure two characteristics of
a test: the family of its primary oracle, which says what the test checks,
and the artifact that states its property, which says whether the program's
documentation or its code states the property.

\par\textbf{Setup.}
\label{sec:rq2-setup}
The population is the \rqtwoTests{} property-based test functions of this
configuration: the \draftnum{2,352} tests of RQ1 (Table~\ref{tab:rq1a-cost})
and \draftnum{270} tests on 20 further PyTorch focal points of RQ4
(Section~\ref{sec:rq4}).
We sample \rqtwoSample{} tests, 50 per target. We visit a target's focal
points in a seeded random order and take the first test of each, then the
second test of each, and so on until we have 50, so no test is drawn twice.
This draw underrepresents a focal point with many tests, so every share
weights a sampled test by the number of tests of its focal point divided by
the number we sampled from it~\cite{cochran1977sampling}. A target's weights
thus sum to its number of tests, and a share over all tests weights each
target by its tests. For the 95\% confidence interval of a share, we use the
normal interval with Kish's effective sample size $(\sum w)^2/\sum w^2$ over
the weights $w$ of the sampled tests~\cite{kish1965survey}, which treats the
sampled tests as independent draws. The interval is at most
$\pm$\rqtwoCIDomain\% within a domain and $\pm$\rqtwoCITotal\% over all
tests, so a difference between two domains is significant only when it
exceeds \rqtwoCIContrast\ points.

\emph{Labels.} The \emph{primary oracle} of a test checks its intended
property, as opposed to checks such as the type of an output the harness
expects. Its family is contract, differential, or metamorphic, which
Section~\ref{sec:domains-chars} defines after Barr
et~al.~\cite{barr2015oracle}. Two rarer families of the same survey also occur:
\emph{algebraic} (an operation followed by its inverse returns the original
value) and \emph{implicit} (the program does not crash). The
\emph{primary source} of a test's property is the artifact that states it
most directly: documentation (manuals and docstrings), code (the code under
test and related code, such as a caller), the project's tests, or another
artifact, such as a standard or a reference implementation.
When a test has several checks, its primary oracle is the one that the test's
name and docstring describe, or else its last assertion on the program's
output. The cronie test of Figure~\ref{fig:dom-sys}, for example, has
a contract as its primary oracle, because the expected times come from a model
that the test itself computes, not from another implementation.

\emph{Methods.} An LLM agent (Claude Opus~5.5) labels the \rqtwoSample{}
tests. A second agent from another vendor (Gemini~3.1 Pro) labels an audit
sample of 100, a seeded random draw of ten tests per target, without seeing
the first agent's labels. An author decides each disagreement. For
the oracle, an agent picks one of ten kinds of check from a closed list.
A script maps the kind to its family, so no labeler chooses a
family directly; the artifact lists the ten kinds and their families. The two agents agree on the family for 82\% of the audit sample
($\kappa$ = 0.72). For the source, a label names the artifact, the location,
and the passage that states the property. A script finds that passage in the
writer's transcript for 473 of the \rqtwoWithSource\ tests whose label names
an artifact; the other 5 passages appear only in the test itself, and the
remaining \rqtwoNoSource\ tests name no artifact. The agents agree on the
source class for 75\% ($\kappa$ = 0.49). In 15 of the 25 disagreements the
second agent finds no artifact, and the author sides with the first agent in
21 and with the second in 4. The agreement shows that the two LLMs label consistently, not that the
labels are correct, and the confidence intervals above cover sampling error
only, not errors that both LLMs make.

\par
\begin{figure}[t]
\centering
\begin{subfigure}[t]{0.55\linewidth}
\centering
\includegraphics[width=\linewidth]{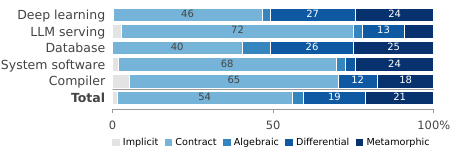}
\caption{Oracle families}
\label{fig:rq2a-family}
\end{subfigure}\hfill
\begin{subfigure}[t]{0.426\linewidth}
\centering
\includegraphics[width=\linewidth]{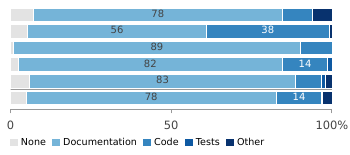}
\caption{Primary sources}
\label{fig:rq2b-primary}
\end{subfigure}
\caption{Oracle families (a) and property sources (b), in weighted \% of tests.}
\label{fig:rq2ab}
\end{figure}

\textbf{Results.}~Figure~\ref{fig:rq2ab} shows the families and the
sources per domain.
Contract oracles are the most common family in every
domain, \rqtwoContractTotal\% of all tests, from \rqtwoContractDB\% in the
databases to \rqtwoContractLLM\% in LLM serving. Of all tests,
\rqtwoContractInvariant\% check an invariant of one output,
\rqtwoContractRejection\% check that the program rejects an invalid input,
and \rqtwoContractDocumented\% check a documented value.
The other \rqtwoNonContractTotal\% of the tests have a primary oracle of
another family: metamorphic
oracles relate two runs (\rqtwoMetamorphicTotal\%), differential oracles
compare with another implementation (\rqtwoDifferentialTotal\%), algebraic
oracles check a round trip (\rqtwoAlgebraicTotal\%), and implicit oracles
check only that the program does not crash (\rqtwoImplicitTotal\%). A test
can also check more than its primary oracle, and Figure~\ref{fig:rq2ab} counts
only the primary one. The metamorphic and differential oracles together are the tests that compare a
run with another run or implementation, \rqtwoCompareTotal\% of all tests.
They are fewest in LLM serving (\rqtwoCompareLLM\%) and most in the deep
learning libraries (\rqtwoCompareDL\%) and the databases
(\rqtwoCompareDB\%). What they compare with depends on the domain. In the
deep learning libraries, tests compare with a reference library such as NumPy
(\rqtwoDifferentialDL\% differential). In system software, they check
equivalences that the manual states (\rqtwoMetamorphicSYS\% metamorphic), such
as between a crontab line starting with
\texttt{@weekly} and one starting with \texttt{0 0 * * 0}.

Documentation states the property of \rqtwoSourceDocumentationTotal\% of all
tests: a docstring or text the program prints for \rqtwoDocstring\%, and a
separate file for \rqtwoDocFile\%. Code states \rqtwoSourceCodeTotal\%. The
labels name no artifact for \rqtwoSourceNoneTotal\%: for
\rqtwoNoSourceNotFound\% the labeler found no passage that states the
property, for \rqtwoNoSourceGeneral\% it judged that the writer used general
knowledge, and \rqtwoNoSourceImplicit\% check only that valid inputs do not
crash. LLM serving is the
exception where documentation states \rqtwoSourceDocumentationLLM\% of the
properties and code \rqtwoSourceCodeLLM\%. In the other four domains
documentation states \rqtwoSourceDocumentationDL\% to
\rqtwoSourceDocumentationDB\% and code \rqtwoSourceCodeCC\% to
\rqtwoSourceCodeSYS\%.

\lessonblock{RQ2}{\agenticpbt's tests check what the program documents:
documentation states \rqtwoSourceDocumentationTotal\% of the properties, and
\rqtwoContractTotal\% of the primary oracles check a condition on one
outcome. Most of the rest compare with a second run or implementation: the
deep learning tests compare with a reference library, and the system software
tests check equivalences that the manuals state. The exception is LLM
serving, where the code states \rqtwoSourceCodeLLM\% of the properties.}

\subsection{RQ3: Developer-Confirmed Bugs}
\label{sec:rq3}

\newcommand{\rqthreecapcompiler}{20 minutes}
\newcommand{\rqthreecapserving}{1 hour}
\newcommand{\rqthreefuzz}{one hour}
\newcommand{\rqthreeCutoffBugs}{10}
\newcommand{\rqthreeCutoffFocal}{13}
\label{sec:rq3-setup}
\par\textbf{Setup.}
\emph{Sessions.} We run \agenticpbt with Opus~5, the configuration that RQ1
selects. The target versions are mostly older than RQ1 and RQ2, because we
began the bug hunting campaigns earlier; Table~\ref{tab:bugs} also counts the
\bugsReportedRQone{} bugs that RQ1's sessions revealed on the versions of
Table~\ref{tab:subjects}. We run one session per focal point,
and we choose the focal points from the target's documentation, its test
files, and the pull requests that fixed bugs. The agent receives only the
focal point's name, such as \texttt{jax.numpy.histogram}. Outside Python it
also receives the harness of RQ1 (Section~\ref{sec:rq1-setup}). Bug hunting
uses a longer session cap than RQ1, so that a session can explore its focal
point further: \rqthreecapcompiler. LLM serving engines are capped at \rqthreecapserving\
to give the engine time to start. The agent reads a shallow checkout of the
pinned commit, with no Git history, in a container whose only network access
is the model's API. These controls block online bug reports and the target's
history. The one attempt to fetch a web page was denied.

\emph{Fuzzing and triage.} After generation, we fuzz every test without an
LLM for \rqthreefuzz\ per focal point on one CPU core. A failure is a test
that violates its oracle or crashes. We deduplicate crashes by their top two
stack frames and other failures by test and exception type. An LLM agent
discards false positives and merges the rest into distinct \emph{candidate
bugs}.
We read a candidate ourselves, reproduce it on the newest release of
the target (on the compilers, the development head), and report it with a
standalone reproducer. A reported candidate is a \emph{bug}, which developers
confirm, reject, or leave pending. A confirmed bug is a \emph{duplicate} if
someone outside this project reported it before a session found it, and
\emph{new} otherwise; new bugs are fixed or await a fix. As in
Section~\ref{sec:intro}, a previously reported bug counts only if its first
public report postdates the training cutoff of the model that found it. When
Opus~5 finds a bug that was first reported before its cutoff (May 2026), we
rerun the focal point in an isolated session with Opus~4.8, whose cutoff is
January 2026. We count the bug only if Opus~4.8 finds it again and the report
postdates January 2026, and otherwise we set the bug aside.

\emph{Measures.} We report new confirmed bugs per focal point, leaving
duplicates out to measure the ability to find new bugs, and the
\emph{rejection rate} per domain, the share of reported bugs that developers
rejected. Candidates we have not read or reported, and pending reports, make the
confirmed rate a lower bound. Because the focal points' sources and the
targets' versions differ by target, the rates describe each domain's run and
do not rank the domains against one another.

\input{tables/bug_table}

\par\textbf{Bugs.}~Table~\ref{tab:bugs} shows the bugs we reported per domain
and target, by status: confirmed, rejected, or pending. It splits the
confirmed ones into those awaiting a fix, those fixed, and the duplicates. We
have reported \bugsReported\ bugs on \bugsSystems\ targets:
\bugsConfirmed\ are confirmed, \bugsPending\ pending, and \bugsRejected\
rejected. Of the confirmed bugs, \bugsDuplicate\ are duplicates, which leaves
\the\numexpr\bugsReported-\bugsDuplicate\relax\ reports not known to
duplicate an earlier one; of these, maintainers confirmed
\the\numexpr\bugsConfirmed-\bugsDuplicate\relax\ new bugs, of which
\bugsFixed\ are fixed. Only \bugsCrash\ of the
\bugsReported\ reported bugs are crashes, uncaught exceptions, or hangs. The
other \bugsFunctional\ are wrong results, which a test detects only through
its oracle.
Our experiments cover \bugsFocal\ focal points in total: \bugsFocalDL\ in
the deep learning libraries, \bugsFocalServing\ in LLM serving,
\bugsFocalDatabase\ in the databases, \bugsFocalSystem\ in system software,
and \bugsFocalCompiler\ in the compilers. We read and
reported \bugsPerFocalReported\ (\bugsReportedFocal/\bugsFocal) bugs per
focal point, and the developers have confirmed \bugsPerFocalNew\
(\bugsConfirmedNewFocal/\bugsFocal) new bugs per focal point, or
\bugsPerFocal\ (\bugsConfirmedFocal/\bugsFocal) counting the duplicates.
In LLM serving, \bugsDuplicateServing\ of the \bugsConfirmedServing\ confirmed
bugs are duplicates, which leaves \bugsPerFocalNewServing\
(\bugsConfirmedNewFocalServing/\bugsFocalServing) new bugs per focal point,
the second-lowest rate after the databases' \bugsPerFocalNewDatabase\
(\bugsConfirmedNewFocalDatabase/\bugsFocalDatabase); system software has the
highest, \bugsPerFocalNewSystem\
(\bugsConfirmedNewFocalSystem/\bugsFocalSystem).
The developers rejected \bugsRejectRate\%
(\bugsRejected/\bugsReported) of the reports:
\bugsRejectRateCompiler\% (\bugsRejectedCompiler/\bugsReportedCompiler) in
the compilers, \bugsRejectRateDL\% (\bugsRejectedDL/\bugsReportedDL) in the
deep learning libraries, \bugsRejectRateServing\%
(\bugsRejectedServing/\bugsReportedServing) in LLM serving, and none in the
databases and system software. In the databases, \bugsPendingDatabase\ of \bugsReportedDatabase\ reports
are still pending (\bugsPendingPostgreSQL\ of PostgreSQL's
\bugsReportedPostgreSQL), so their \bugsPerFocalNewDatabase\ new confirmed
bugs per focal point is a lower bound, which rises to
\bugsPerFocalMaxDatabase\ if maintainers confirm every pending report as new.

\begin{figure}[t]
\centering
\begin{subfigure}[t]{0.48\linewidth}
\begin{tcblisting}{codebox, equal height group=rq3ex,
  minted language=js, minted options={fontsize=\scriptsize}}
RETURN SUBSTITUTE("", "#", "replacement")
// [ null ], with a warning
// expected: [ "" ]
\end{tcblisting}
\vspace{-0.6em}
\caption{ArangoDB's \texttt{SUBSTITUTE}.}
\label{fig:rq3-arangodb}
\end{subfigure}\hfill
\begin{subfigure}[t]{0.48\linewidth}
\begin{tcblisting}{codebox, equal height group=rq3ex,
  minted language=c, minted options={fontsize=\scriptsize}}
__attribute__((vectorcall))
int f(long double a) { return (int)a; }
// Clang crashes: assertion failure
\end{tcblisting}
\vspace{-0.6em}
\caption{Clang's \texttt{vectorcall} attribute.}
\label{fig:rq3-clang}
\end{subfigure}
\vspace{-0.6em}
\caption{Two bugs that \agenticpbt found, as minimal reproducers.}
\label{fig:rq3-bugs}
\vspace{-0.8em}
\end{figure}
\par\textbf{Examples.}~Figure~\ref{fig:rq3-bugs} shows two of these bugs as
minimal reproducers. Figure~\ref{fig:rq3-arangodb} shows a bug in
ArangoDB's AQL function \texttt{SUBSTITUTE(value, search, replace)}, which,
as its manual states, replaces ``all occurrences of \texttt{search}'' in
\texttt{value} with \texttt{replace}. The test draws the three strings from lowercase
letters, runs the query through the harness, and requires the result to equal
that of Python's \texttt{str.replace}, without an exception and without a
warning. It fails when
\texttt{value} is empty: \texttt{SUBSTITUTE} returns \texttt{null} with a
warning instead of the empty string. The developers confirmed the bug, merged
a fix the next day, and released it in ArangoDB 3.12.12~\cite{arangodb-23293}.

Figure~\ref{fig:rq3-clang} shows a bug in Clang's \texttt{vectorcall}
attribute, which, as Clang's manual states, ``changes the calling convention
of a function,'' i.e., the way a call passes the function's arguments. The test
draws a return type and up to nine parameter types from scalar, vector, and
structure types. It then builds a program that defines the same function with
and without the attribute. Through the harness, it compiles and runs the
program, and requires that both functions return the same value without
crashing.
It fails when a parameter is a \texttt{long double}: Clang crashes with
an assertion failure during code generation instead of compiling the function.
An LLVM developer labeled our report as confirmed~\cite{llvm-227448}.

\lessonblock{RQ3}{Developers confirmed bugs in all five domains. Of the
\bugsReported\ bugs we reported, \bugsConfirmed\ are confirmed,
\the\numexpr\bugsConfirmed-\bugsDuplicate\relax\ of them new and
\bugsFixed\ fixed, and \bugsRejected\ rejected (\bugsRejectRate\%). Only \bugsCrash\ are
crashes, uncaught exceptions, or hangs; the rest are wrong results that only
an oracle detects. New confirmed bugs per focal point range from
\bugsPerFocalNewDatabase\ in the databases and \bugsPerFocalNewServing\ in
LLM serving, where most confirmed bugs were already known, to
\bugsPerFocalNewSystem\ in system software.}

\FloatBarrier  %
\subsection{RQ4: Comparison with Domain-Specific Tools}
\label{sec:rq4}
\label{sec:rq1b} %

RQ4 asks how much \agenticpbt's tests and the fuzzers built for three of
our domains cover, and which crashes each finds that the other does not. The fuzzers' oracles
differ from a property-based test's, so other failures are not comparable: MetaMut and GrayC detect
compiler crashes, Centaur and FlashFuzz PyTorch crashes, and SQLancer
crashes and one fixed relation between queries.

\par\textbf{Setup.}~\emph{Tools and targets.} We compare with
MetaMut~\cite{ou2024metamut} and GrayC~\cite{evenmendoza2023grayc} on GCC and
Clang, with SQLancer~\cite{rigger2020tlp} on PostgreSQL, and with
Centaur~\cite{naziri2026centaur} and FlashFuzz~\cite{qin2026effective} on
PyTorch. We omit JAX, which the evaluated versions of Centaur and FlashFuzz
do not support, and ArangoDB, whose support SQLancer has
removed~\cite{sqlancer-nosql-removal}. \drafttext{Each tool runs in its
default configuration, and the artifact lists the versions.} MetaMut and
GrayC start from the seed corpora of their releases.
\emph{Budget.} \agenticpbt uses Opus~5 and RQ1's generation setup. Each tool then runs once for \drafttext{12 hours on 20 CPU cores} on the
instrumented builds of RQ1; the 12 hours cover fuzzing only, and
\agenticpbt's test generation (Table~\ref{tab:rq1a-cost}) precedes them. \agenticpbt's tests share the budget equally and run one after another on
each core, so its early coverage is lower than in RQ1
(Figure~\ref{fig:rq1a-cov}).
\emph{Coverage.} On PyTorch, all three tools test the same \numfocalpt\ APIs,
sampled from those that both fuzzers support, and \drafttext{we measure each
tool's branch coverage only during calls to these APIs, as Centaur's
evaluation does~\cite{naziri2026centaur}}. On the other targets, we measure
the coverage of the whole program.
\emph{Crashes.} We replay the inputs that \agenticpbt, MetaMut, and GrayC
saved on the evaluated compiler builds, replay FlashFuzz's inputs on PyTorch,
and check the child process's signal for \agenticpbt's PyTorch crash. We
count the crashes that reproduce, whether or not they were reported before,
and deduplicate them per compiler and tool by the first two function names
in the stack trace.

\begin{figure}[t]
\centering
\includegraphics[width=\linewidth]{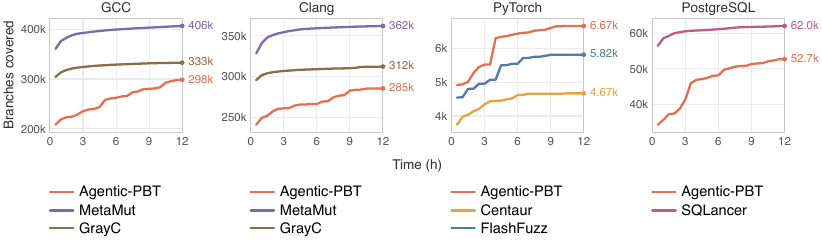}
\caption{\drafttext{Branch coverage over 12 hours, one run per tool.}}
\label{fig:rq1b-cov}
\end{figure}

\begin{wraptable}{r}{0.4\linewidth}
\vspace{-1.2em}
\centering
\scriptsize
\caption{Reproduced crashes per tool, and those shared with \agenticpbt{} (--: not run on the target).}
\label{tab:rq4-crashes}
\setlength{\tabcolsep}{2.5pt}\renewcommand{\arraystretch}{0.8}
\setlength{\aboverulesep}{0.2ex}\setlength{\belowrulesep}{0.3ex}
\begin{tabular}{@{}lcccc@{}}
\toprule
\textbf{Tool} & \textbf{GCC} & \textbf{Clang} & \textbf{PyTorch} & \textbf{PostgreSQL} \\
\midrule
\agenticpbt & \draftnum{2} & \draftnum{2} & \draftnum{1} & \draftnum{0} \\
MetaMut & \draftnum{7} & \draftnum{16} & -- & -- \\
GrayC & \draftnum{1} & \draftnum{3} & -- & -- \\
FlashFuzz & -- & -- & \draftnum{1} & -- \\
Centaur & -- & -- & \draftnum{0} & -- \\
SQLancer & -- & -- & -- & \draftnum{0} \\
\midrule
Shared & \draftnum{0} & \draftnum{0} & \draftnum{1} & \draftnum{0} \\
\bottomrule
\end{tabular}
\vspace{-1em}
\end{wraptable}
\par\drafttext{\textbf{Coverage.}~Figure~\ref{fig:rq1b-cov} shows the
coverage over the 12 hours. MetaMut covers the most branches on GCC and
Clang; GrayC reaches \draftnum{82--86\%} of its coverage and \agenticpbt
\draftnum{73--79\%}. SQLancer leads on PostgreSQL with \draftnum{61,984}
branches, of which \agenticpbt reaches \draftnum{85\%}. \agenticpbt leads on
PyTorch with \draftnum{6,670} branches, \draftnum{1.1} times FlashFuzz's
coverage and \draftnum{1.4} times Centaur's. On the compilers and PostgreSQL, \agenticpbt's tests target 30 focal
points per program while the fuzzers exercise the whole program, so
whole-program coverage favors the fuzzers; on PyTorch, all three tools are
confined to the same 50 APIs.}

\par\textbf{Crashes.}~Table~\ref{tab:rq4-crashes} gives the reproduced
crashes per tool and target. On the compilers, MetaMut finds the most,
\draftnum{7} on GCC and \draftnum{16} on Clang, against \agenticpbt's
\draftnum{2} and \draftnum{2}, and no crash signature of \agenticpbt matches one of MetaMut's or GrayC's.
\agenticpbt's tests do not seek crashes; its four are property tests that
crashed the compiler instead of failing. On PyTorch, \agenticpbt and FlashFuzz find the
same crash, in PyTorch's \texttt{GRUCell} module; Centaur finds none. No tool crashes
PostgreSQL. \agenticpbt's two Clang crashes include the \texttt{vectorcall}
crash of RQ3 (Figure~\ref{fig:rq3-clang}), whose test an RQ1 session wrote; the counts also include crashes reported before our runs, such as
GrayC's crash on huge by-value arguments~\cite{llvm-118207}.

\lessonblock{RQ4}{\drafttext{The domain-specific fuzzers cover more of GCC,
Clang, and PostgreSQL, where \agenticpbt reaches \draftnum{73--85\%} of the
leader's coverage; \agenticpbt covers more of the sampled PyTorch APIs.
MetaMut and GrayC find \draftnum{23} and \draftnum{4} compiler crashes to
\agenticpbt's \draftnum{4}, and none of \agenticpbt's matches one of theirs.
The comparison covers crashes, the one failure both sides detect; the
oracle-detected bugs of RQ3 lie outside it.}}

\FloatBarrier  %

%% file: tables/bug_table.tex
\begin{table}[t]
\centering
\footnotesize
\setlength{\tabcolsep}{2.25pt}
\renewcommand{\arraystretch}{0.9}
\setlength{\aboverulesep}{0.2ex}\setlength{\belowrulesep}{0.3ex}
\caption{Bugs \agenticpbt{} found and we reported so far, by domain, target,
  and release.}
\label{tab:bugs}
\begin{tabular}{@{}lllrrrrrrr@{}}
\toprule
\multirow{2}{*}{\textbf{Domain}} & \multirow{2}{*}{\textbf{Target}} & \multirow{2}{*}{\textbf{Version}} & & \multicolumn{3}{c}{\textbf{Status}} & \multicolumn{3}{c}{\textbf{Of the confirmed}} \\
\cmidrule(lr){5-7}\cmidrule(lr){8-10}
 &  &  & \textbf{Reported} & \textbf{Confirmed} & \textbf{Rejected} & \textbf{Pending} & \textbf{Awaiting} & \textbf{Fixed} & \textbf{Duplicate} \\
\midrule
\rowcolor{black!8}
 & JAX & 0.10.1 & 38 & 28 & 1 & 9 & 14 & 11 & 3 \\
\rowcolor{black!8}
\multirow{-2}{*}{Deep learning} & PyTorch & 2.12.0 & 12 & 8 & - & 4 & 4 & 1 & 3 \\
 & vLLM & 0.28.0 & 33 & 27 & - & 6 & 4 & 3 & 20 \\
\multirow{-2}{*}{LLM serving} & SGLang & 0.5.18 & 33 & 17 & 1 & 15 & - & 1 & 16 \\
\rowcolor{black!8}
 & PostgreSQL & 18.6 & 9 & 1 & - & 8 & - & 1 & - \\
\rowcolor{black!8}
\multirow{-2}{*}{Database} & ArangoDB & 3.12.4 & 23 & 5 & - & 18 & - & 4 & 1 \\
 & OpenSSH & 8.9p1 & 5 & 1 & - & 4 & - & 1 & - \\
\multirow{-2}{*}{System software} & cronie & 1.7.2 & 13 & 13 & - & - & 12 & 1 & - \\
\rowcolor{black!8}
 & GCC & 17 (33097f2) & 5 & 3 & 1 & 1 & 3 & - & - \\
\rowcolor{black!8}
\multirow{-2}{*}{Compiler} & Clang & 22.1.8 & 9 & 6 & - & 3 & 5 & - & 1 \\
\midrule
Total &  &  & 180 & 109 & 3 & 68 & \cellcolor{black!8}\textbf{42} & \cellcolor{black!8}\textbf{23} & 44 \\
\bottomrule
\end{tabular}
\end{table}

%% file: discussion.tex
\section{Discussion}
\label{sec:discussion}

\subsection{Lessons Learned}
\label{sec:lessons}

\drafttext{We draw three lessons from our results: two for building and
evaluating PBT writers, and one on how such writers relate to the fuzzers
that already test these domains.}

\drafttext{\textbf{(1) Low observability hurts PBT writers.} \emph{Observability} is how
much of a program's behavior its outputs reveal~\cite{freedman1991testability}.
A property can be documented and still be hard to check, because the program
does not show the behavior that the property concerns. This affects system-level tests most, which see only what the program
prints. crontab(5) states which lines of a
crontab are environment settings and which are jobs, but cronie's output does
not say how it read a given line. An \agenticpbt test works around this with cronie's limit of 1,000
environment settings: it writes 1,000 settings and then the line under test,
so cronie reports an error if it reads that line as a setting and accepts the
file if it reads it as a job. The test thus depends on knowledge beyond the
property, which the LLM supplied. A writer facing
low observability needs either the knowledge, from the LLM or from the user
through a prompt that asks for it, or a way to raise observability, for
example probes that expose the program's internal decisions, neither of
which we tried.}

\drafttext{\textbf{(2) Evaluate a writer's reports as well as its final tests.} While it writes its tests, a writer can find a failure and then stop
generating the inputs that reveal it. On \texttt{torch.fft.ifftshift}, the
session reported a failure for calls with no dimensions to rearrange, but its
final test no longer generates such calls. \agenticpbt at least keeps such failures in its reports. Evaluations should
therefore
include a writer's reports, and writers should record the inputs they
exclude.}

\drafttext{\textbf{(3) Fuzzers supply inputs, and LLM-based writers supply
properties.} The documentation states the property of \rqtwoSourceDocumentationTotal\% of \agenticpbt's tests, and
source code that of \rqtwoSourceCodeTotal\% (Section~\ref{sec:rq2}). Only \bugsCrash{} of the \bugsReported{} bugs are crashes or hangs; the
rest are wrong results that only an oracle detects (Section~\ref{sec:rq3}). The domain-specific tools of RQ4 cover more
of the compilers and of PostgreSQL than \agenticpbt, but their oracles check crashes and hangs or, in SQLancer's case, one fixed
relation between a query and its partitions; none checks the documented
behavior of a feature. Given
properties that an LLM mines from documentation and code, a fuzzer could also
check what each feature should compute.}

\subsection{Threats to Validity} 

\emph{Internal validity.} Each writer generates its tests once per
configuration; the repeated fuzzing measures the variability of those tests,
not of their generation. RQ4 runs each tool only once. We wrote the
harnesses, which could favor one
writer, but both use the same harness, which only runs the program and
returns its output. Humans judge RQ1's sampled failures, but an LLM agent
triages RQ3's failures and LLM agents label RQ2's tests; a second vendor's
agent relabels an audit sample and an author decides each disagreement, and the labels with low agreement remain a threat. The triage agent of RQ3
and the primary labeler of RQ2 are Claude models, from the vendor of the
writer's main LLM; the Gemini audit mitigates this for RQ2, and for RQ3 we read every candidate we report, although we do not review
the failures the triage discards. 
A bug may be in a model's
training data, so we exclude bugs reported before the model's cutoff and
recheck earlier ones with a model whose cutoff is earlier.
\emph{Construct validity.} Branch coverage of a whole target includes code that
every test reaches, such as start-up code, and coverage does not measure fault
finding. Unique failures depend on how we deduplicate them, by error message or
stack trace, which can merge distinct bugs or split one. Confirmed bugs depend on maintainers, and pending reports make our rates
lower bounds.
\emph{External validity.} We study two programs in each of five domains, two
writers, and two LLMs, so our results may not hold for other programs, domains, writers, or
models. We sample 30 focal points per target, and on PyTorch RQ4 samples
only the APIs that Centaur and FlashFuzz support. All tests use Hypothesis,
on selected versions of each target.

%% file: related.tex
\section{Related Work}
\label{sec:related}

\textbf{LLM-based PBT writers.} \proptestai~\cite{vikram2024pbt} and
\agenticpbt~\cite{maaz2025agentic} (Section~\ref{sec:bg-llm}) are the two
general-purpose writers we know of, and both were evaluated only on Python
libraries. He et al.~\cite{he2025pbtcodegen} use LLM-written PBTs to validate and refine
LLM-generated code, and Gentoo~\cite{vikram2026gentoo} has a coding agent write
input generators for Java fuzz targets, without oracles. 

\textbf{Mining properties and oracles.} Before LLMs, each tool mined one kind
of property from one source. Daikon~\cite{ernst2001dynamically} infers likely
invariants from execution traces; Toradocu~\cite{goffi2016toradocu},
MeMo~\cite{blasi2021memo}, and CallMeMaybe~\cite{blasi2022callmemaybe} extract
exception, metamorphic, and temporal oracles from Javadoc comments;
MR-Scout~\cite{xu2024mrscout} mines metamorphic relations from existing tests;
and DeepREL~\cite{deng2022fuzzing} infers relations between the APIs of deep
learning libraries. TOGA~\cite{dinella2022toga} and ChatAssert~\cite{hayet2025chatassert}
generate assertions for a given test with a neural model or an LLM, and
Endres et al.~\cite{endres2024nl2postcond} have an LLM turn natural-language
intent into postconditions. An LLM-based PBT writer instead produces the
generator, the property, and the oracle together, from whatever artifact
states the property.

\textbf{Test generation and fuzzing.} Property-based testing has been
joined with coverage-guided fuzzing before LLMs: JQF and
Zest~\cite{padhye2019jqf,padhye2019zest}, FuzzChick~\cite{lampropoulos2019fuzzchick},
and RLCheck~\cite{reddy2020rlcheck} steer a generator toward valid and
diverse inputs with coverage feedback or reinforcement learning, but the
property and the generator come from the user. LLMs also write unit
tests~\cite{schafer2024testpilot,lemieux2023codamosa} and fuzz inputs, either
for one domain, as TitanFuzz~\cite{deng2023titanfuzz} does for deep learning
libraries, or for many languages at once, as Fuzz4All~\cite{xia2024fuzz4all} does;
WhiteFox~\cite{yang2024whitefox}, KernelGPT~\cite{yang2025kernelgpt}, and
ChatAFL~\cite{meng2024chatafl} have an LLM read a compiler's, a kernel's, or
a protocol's source or specification to steer a fuzzer's inputs. Unit tests check one input each, fuzzers detect mostly crashes, and
LLM-based test generators can discard the tests that reveal
bugs~\cite{mathews2024design}.

\textbf{Domain-specific testing.} The system-level domains of our study have
mature testing tools: Csmith~\cite{yang2011csmith}, EMI~\cite{le2014emi},
YARPGen~\cite{livinskii2020yarpgen}, MetaMut~\cite{ou2024metamut}, and
GrayC~\cite{evenmendoza2023grayc} for compilers, and
RAGS~\cite{slutz1998rags}, PQS~\cite{rigger2020pqs},
NoREC~\cite{rigger2020norec}, TLP~\cite{rigger2020tlp}, and
Squirrel~\cite{zhong2020squirrel} for databases; deep learning libraries have
Centaur~\cite{naziri2026centaur} and FlashFuzz~\cite{qin2026effective}, and
testing LLM serving engines is just
beginning~\cite{liu2026firstlook,zhao2026grief}. RQ4 compares \agenticpbt with five of them.

%% file: conclusion.tex
\section{Conclusion}
\label{conclusion}

We asked how far LLM-based PBT writers, built and evaluated on Python
libraries, carry over to compilers, databases, system software, and LLM serving
engines. They carry over, with two limits. \agenticpbt wrote property-based tests in all
five domains, took most of their properties from the documentation, and
revealed \keynum{\the\numexpr\bugsConfirmed-\bugsDuplicate\relax{}} new bugs
that maintainers confirmed in all ten programs; most of the bugs we
reported are wrong results that only an oracle detects. It covered more
than \proptestai on seven of the ten programs with Opus~5 and on all ten with
GLM-5.3-Flash. The first limit is coverage: the generators reach 73 to 85\%
of a mature fuzzer's coverage on the compilers and PostgreSQL, where MetaMut
finds \draftnum{23} compiler crashes against \agenticpbt's \draftnum{4}, none
in common; on PyTorch, \agenticpbt covers more than its fuzzers. The second
is observability: some documented properties are hard to check because the
program's output does not reveal the behavior they concern. The coverage gap
suggests a remedy. The fuzzers check for crashes or, in SQLancer's case, for
one fixed relation between queries, while the writer supplies the documented
property; a fuzzer's generator steered toward a property's focal point could
reach the inputs that the writer's generator does not.

%% file: data.tex
\section*{Data Availability}
\label{sec:data}

Our artifacts are publicly available at \RepoURL.